\documentclass[onecolumn,nofootinbib,superscriptaddress,notitlepage,amsmath,amssymb,fleqn,aps,prd,preprintnumbers,10pt]{revtex4-1}
\usepackage{xcolor,graphicx,appendix,xspace,hyperref}
\usepackage[export]{adjustbox}
\usepackage[all]{hypcap}
\usepackage[normalem]{ulem}
\usepackage{booktabs}

\newcommand{\powhegbox}{{\sc Powheg\,Box}\xspace}

\newcommand{\pythia}{{\sc Pythia}\xspace}
\newcommand{\dire}{{\sc Dire}\xspace}
\newcommand{\vincia}{{\sc Vincia}\xspace}
\newcommand{\herwig}{{\sc Herwig}\xspace}
\newcommand{\sherpa}{{\sc Sherpa}\xspace}
\newcommand{\nnlojet}{{\sc NNLOJet}\xspace}
\newcommand{\minlo}{MiNLO\xspace}
\newcommand{\minnlops}{MiNNLO$_\text{PS}$\xspace}
\newcommand{\rivet}{{\sc Rivet}\xspace}
\newcommand{\fastjet}{{\sc FastJet}\xspace}
\newcommand{\zenodoplotref}{\href{https://doi.org/10.5281/zenodo.17015614}{https://doi.org/10.5281/zenodo.17015614}\xspace}

\newcommand{\gev}{\mathrm{GeV}}

\definecolor{comment}{rgb}{0,0.3,0}
\usepackage{listings}
\newcommand{\affFermilab}{Fermi National Accelerator Laboratory, Batavia, IL 60510, USA}
\newcommand{\affShandong}{School of Physics, Shandong University, Jinan, Shandong 250100, China}
\newcommand{\affMSU}{Department of Physics and Astronomy, Michigan State University, MI 48824, USA}
\newcommand{\affBuffalo}{University at Buffalo, The State University of New York, Buffalo, NY 14260, USA}
\newcommand{\affMuenster}{Institut f{\"u}r Theoretische Physik, Universit{\"a}t M{\"u}nster, D-48149 M{\"u}nster, Germany}
\newcommand{\affGoettingen}{Institut f{\"u}r Theoretische Physik, Georg-August-Universit{\"a}t G{\"o}ttingen, D-37077 G{\"o}ttingen, Germany}
\newcommand{\affNoertheastern}{Northeastern University, Boston, MA 02115, USA}
\newcommand{\affCERN}{Theoretical Physics Department, CERN, CH-1211 Geneva, Switzerland}
\newcommand{\affCERNexp}{Experimental Physics Department, CERN, CH-1211 Geneva, Switzerland}

\hypersetup{hidelinks,
  pdfauthor={Xuan Chen,Silvia Ferrario Ravasio,Yacine Haddad,Stefan Hoeche,Joey Huston,Tomas Jezo,Jia-Sheng Liu,Christian Preuss,Ahmed Tarek,Jan Winter},
  pdftitle={Theory uncertainties of the irreducible background to VBF Higgs production},
  pdfkeywords={QCD, Perturbation Theory}
}
\begin{document}
\preprint{CERN-TH-2025-176, FERMILAB-PUB-25-0626-T, MS-TP-25-24, MCNET-25-23}
\title{Theory uncertainties of the irreducible background to VBF Higgs production}
\author{Xuan~Chen}\affiliation{\affShandong}
\author{Silvia~Ferrario Ravasio}\affiliation{\affCERN}
\author{Yacine Haddad}\affiliation{\affNoertheastern}
\author{Stefan~H{\"o}che}\affiliation{\affFermilab}
\author{Joey~Huston}\affiliation{\affMSU}
\author{Tom\'a\v{s}~Je{\v z}o}\affiliation{\affMuenster}
\author{Jia-Sheng~Liu}\affiliation{\affShandong}
\author{Christian~T.~Preuss}\affiliation{\affGoettingen}
\author{Ahmed~Tarek}\affiliation{\affCERNexp}
\author{Jan~Winter}\affiliation{\affBuffalo}

\begin{abstract}
Higgs boson production through gluon fusion in association with two jets is an irreducible background
to Higgs boson production through vector boson fusion, one of the most important channels for
analyzing and understanding the Higgs boson properties at the Large Hadron Collider.
Despite a range of available simulation tools, precise predictions for the corresponding
final states are notoriously hard to achieve. Using state-of-the-art fixed-order calculations as
the baseline for a comparison, we perform a detailed study of similarities and differences in
existing event generators. We provide consistent setups for the simulations that can be used
to obtain identical parametric precision in various programs used by experiments. 
We find that NLO calculations for the two-jet final state are essential to achieve
reliable predictions.
\end{abstract}

\maketitle

\section{Introduction}
The discovery of the Higgs boson~\cite{ATLAS:2012yve,CMS:2022dwd} by the ATLAS and CMS
experiments at the Large Hadron Collider (LHC) has ushered in a new era in elementary particle physics.
The so far excellent agreement of the boson's properties with Standard Model expectations
leaves as many questions open as it answers, and calls for a more detailed investigation of this unique particle. 
One of the most promising Higgs production modes for in-depth studies is weak vector boson fusion,
conventionally referred to as VBF. This process has been analyzed by both ATLAS and CMS for a number
of Higgs boson decay channels at a number of center-of-mass energies~\cite{ATLAS:2020cvh,ATLAS:2020bhl,ATLAS:2022ooq,ATLAS:2022vkf,ATLAS:2025pxp,CMS:2012qbp, CMS:2024ddc, CMS:2021wlt, CMS:2021kom, CMS:2017zyp, CMS:2019jdw,CMS:2022qva}.  
High-statistics measurements of VBF allow for a precise determination of the couplings
of the Higgs boson to $W$ and $Z$ bosons, as well as improved limits in searches for possible new physics.
The presence of two (or more) jets in the final state, characterized by a large rapidity separation
{and} a large dijet invariant mass, can be used to enhance the VBF sub-process over
the dominant production mechanism, which is gluon-gluon fusion (ggF). However, the
final-state configurations in gluon fusion with two (or more) jets can still be identical to
the ones in VBF, turning ggF from a signal into an irreducible background. The full potential
of the LHC for VBF Higgs boson measurements can therefore only be realized, once gluon fusion
Higgs production with two additional jets is theoretically understood to the highest achievable precision. 

Matrix element plus parton shower matching~\cite{Frixione:2002ik,Nason:2004rx,Frixione:2007vw,
  Alioli:2010xd,Hoche:2010pf,Hoeche:2011fd,Alwall:2014hca,Jadach:2015mza,Nason:2021xke,Hoche:2025gsb}
and merging~\cite{Andre:1997vh,Catani:2001cc,Lonnblad:2001iq,Mangano:2001xp,Krauss:2002up,
  Lavesson:2007uu,Hoeche:2009rj,Hamilton:2009ne,Lonnblad:2011xx,Lonnblad:2012ng,
  Platzer:2012bs,Hoche:2019ncc,Lavesson:2008ah,Hoeche:2012yf,Frederix:2012ps,Lonnblad:2012ix,
  Hoeche:2014lxa}
techniques are an essential theory tool to achieve this goal, and are used
in the extraction of the Higgs boson cross sections from experimental measurements.
They are embedded in complete frameworks for the production of the particle-level
final state provided by the general-purpose LHC event generators, \herwig, \pythia, and
\sherpa~\cite{Buckley:2011ms,Campbell:2022qmc}. Many other theory tools make use of
the event generators to provide the implementation of resummation and/or hadronization
in terms of parton showers and fragmentation models.
 The interplay between
perturbative and non-perturbative QCD is a key aspect of Monte Carlo event
generation at the LHC. Simulations typically assume a scale hierarchy,
where the hardest interactions of the event are described by perturbatively computed matrix
elements, while the parton shower adds softer radiation using resummed perturbation theory.
At some point during the QCD evolution, a scale is reached where the transition to the
non-perturbative regime must be implemented. The simulation of this hadronization process
relies on physically motivated models that are tuned to experimental data from current and
past experiments, notably data from LEP, HERA and the LHC. As these data sets are the same
for all event generators, the different hadronization models usually predict fairly similar
results, and their uncertainties tend to be of a similar size~\cite{Hoeche:2011fd,SM:2012sed},
although in cases where particle spectra are resolved in detail, significant differences
can still occur~\cite{Carli:2010cg,Divisova:2025xqp}. On the other hand, one often observes
discrepancies in the simulation of perturbative QCD effects.
A classical example would be if the parton shower settings involved a different form of the
strong running coupling than the hard matrix elements, or if the value of the running coupling
differed at the reference scale.  Such mismatches are due to the attempt to ``tune away''
effects that can arise from missing higher-order corrections. Inconsistent tunes, however,
are not predictions based on solid QCD perturbation theory, and the corresponding
simulation results can no longer be relied upon.

In a previous publication~\cite{Buckley:2021gfw}, some of us focused on a consistent way to simulate
VBF Higgs boson production at both fixed-order (LO, NLO, NNLO) and fixed-order plus parton shower,
finding agreement in both the central predictions and their uncertainties at the level expected
of the perturbative QCD result. These calculations are in line with the observations of
a range of other studies~\cite{Jager:2014vna,Jager:2020hkz,Ballestrero:2018anz,Hoche:2021mkv,Bittrich:2021ztq,Chen:2021phj,vanBeekveld:2023chs,Barone:2025jey}.
In the present paper, we will explore the gluon-gluon fusion background, using
a variety of tools in the same common setup as required to obtain theoretically
consistent results. 

{The studies performed in the present work} match those used by ATLAS and CMS (referenced earlier) as much as possible.
We will show that the nominal setups for the different programs lead to similar predictions
for the kinematic distributions crucial to the determination of the VBF Higgs signal.
As a result, the uncertainties quoted for the ggF backgrounds by ATLAS and CMS may be overstated.
We will compare predictions from approaches that are inclusive in the number of jets
($\text{Higgs} + 0,1,2~\text{jets}$; such as from \sherpa MEPS@LO and MEPS@NLO)
to those from more exclusive approaches, where only the final 2 jet state is produced
(such as \sherpa S-MC@NLO $Hjj$ and \powhegbox + \pythia / \powhegbox + \herwig $Hjj$).
We also calculate the \minnlops prediction for gluon-fusion (ggF) Higgs boson production, in which
the leading jet spectrum is computed at NLO precision, while the second jet is predicted
at LO. This is the framework which is currently used by the ATLAS and CMS experiments.
We compare results at the parton shower level and the hadron level
and discuss similarities and differences. We disentangle uncertainty
estimates as much as possible, in order to provide a better understanding of their sources.
We also test the impact of the merging scheme and scale. We provide a \rivet routine
that can be directly used by the experiments to confirm the correct use of each generator.
Although the ATLAS and CMS determinations of the VBF Higgs cross section are not strictly
cut-based, we will also discuss a cutflow analysis, indicating the results of successive
phase-space restrictions that are used in VBF Higgs analyses, as well as providing comparisons
of what has been termed {\it loose} and {\it tight} VBF selection cuts. 

This manuscript is structured as follows: Section~\ref{sec:setup} introduces the event
generators we use for the study, and explains the setup of the perturbative calculation
and the non-perturbative tunes. It also lists the setups in which the event generators
are typically employed by the experiments. In Sec.~\ref{sec:previous_simulations} we
compare the results from those generators to the predictions typically obtained from
the current LHC simulations. Section~\ref{sec:uncertainties} summarizes our
findings on uncertainties, while Secs.~\ref{sec:higgspt} and~\ref{sec:cutflow}
present some additional phenomenological results. Section~\ref{sec:conclusions}
contains our conclusions and an outlook.

\section{Input parameters and Computational Tools}
\label{sec:setup}
\begin{figure}[t]
    \centering
    \includegraphics[width=0.5\linewidth]{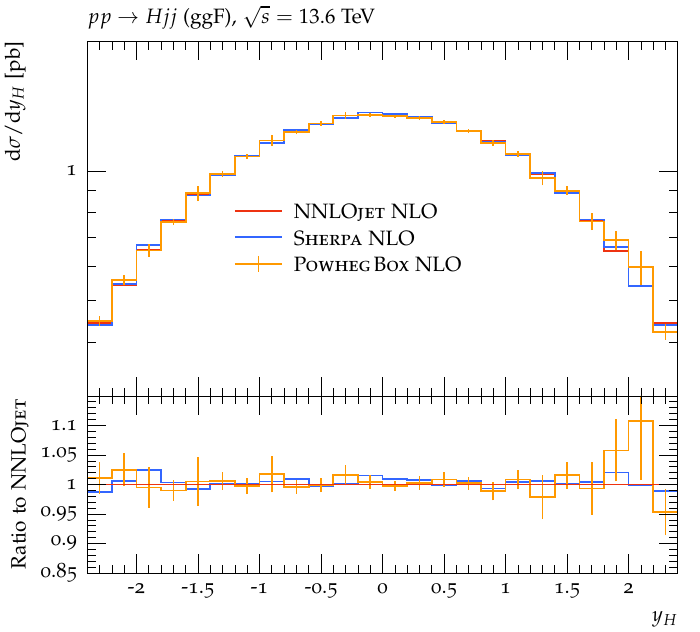}
    \caption{The rapidity distribution of the Higgs at NLO accuracy, computed with \nnlojet (red), \sherpa (blue), and \powhegbox (orange).}
    \label{fig:crossCheckNLO}
\end{figure}

To provide meaningful theoretical guidelines for experimental analyses and simplified
template cross-sections (STXS), we adopt the same jet definition as in ATLAS
measurements~\cite{Buhrer:2019npn,ATLAS:2019jst} and study differential observables
using fiducial bins suggested by STXS stage 1.1~\cite{Berger:2019wnu}. We focus on on-shell
Higgs bosons of $m_{\text{H}}=125$~GeV, produced in the ggF channel with a center-of-mass
energy of 13.6~TeV. The Higgs vacuum expectation value is $v=246.22~\mathrm{GeV}$.
Theoretical uncertainties are estimated by varying QCD renormalization ($\mu_\text{R}$)
and factorization ($\mu_\text{F}$) scales independently by a factor of two around the
central scale of $\mu=H_\text{T}^{\text{parton}}/2$, where
\begin{equation}\label{eq:murf}
   H_\text{T}^{\text{parton}} = \sqrt{m^2_{\text{H}}+p_{\text{T},\text{H}}^2} + \sum_{i \in \text{partons}} p_{\text{T}, i}\;.
\end{equation}
We eliminate the two antipodal combinations $(\mu_\text{R},\mu_\text{F})=(\mu/2,2\mu)$
and $(\mu_\text{R},\mu_\text{F})=(2\mu,\mu/2)$. The parton distribution functions (PDFs)
are defined by the central group of PDF4LHC21\_40 NNLO PDFs~\cite{PDF4LHCWorkingGroup:2022cjn}.
The computational tools used in this study are introduced in the following.
An overview of the perturbative accuracy achieved in each of the tools in the $H$, $H+1j$,
and $H+2j$ phase spaces is given in Tab.~\ref{tab:overviewTools}. We have cross checked
the agreement between \nnlojet, \sherpa, and \powhegbox at NLO. In Fig.~\ref{fig:crossCheckNLO},
we show the rapidity distribution of the Higgs boson as an example.

\begin{table}[t]
    \centering
    \begin{tabular}{l||c|c|c}
    ~ & \multicolumn{3}{|c}{Perturbative Accuracy} \\
    Setup & $H$ & $H+1j$ & $H+2j$ \\ \hline\hline
    \nnlojet & - & NNLO & NLO \\ \hline
    \minnlops~+~\pythia/\herwig & NNLO+PS & NLO+PS & LO+PS \\
    \sherpa MEPS@NLO & NLO+PS & NLO+PS & NLO+PS \\ \hline
    \sherpa MC@NLO & - & - & NLO+PS \\
    \powhegbox~+~\pythia/\herwig & - & - & NLO+PS
    \end{tabular}
    \caption{Comparison of the perturbative accuracy of the different computational tools used in this study in the $H$, $H+1j$, and $H+2j$ fiducial phase spaces.}
    \label{tab:overviewTools}
\end{table}

\subsection{NNLOJet}
\nnlojet\ is a parton level event generator specializing in the computation of NNLO QCD corrections~\cite{NNLOJET:2025rno}. The antenna subtraction formalism~\cite{Gehrmann-DeRidder:2005alt,Gehrmann-DeRidder:2005btv,Daleo:2006xa,Daleo:2009yj,Gehrmann-DeRidder:2012too,Currie:2013vh} is used to regulate IR divergences at each stage of the fixed-order calculations and to provide fully-differential results. For Higgs production in the gluon-fusion channel, \nnlojet\ provides $H+j$ predictions at NNLO~\cite{Chen:2014gva,Chen:2016zka}. This implies that for observables with at least one ($H+j$), two ($H+2j$) and three jets ($H+3j$), the corresponding fixed order QCD accuracy is NNLO, NLO and LO, respectively. In this study, the calculation is performed in the Higgs effective field theory (HEFT) with infinitely large top-quark mass. The relevant Standard Model (SM) top-quark mass effects were studied for $H+2j$ at NLO with full theory approximation on the two-loop matrix elements finding comparable NLO/LO ratios between HEFT and SM for differential observables~\cite{Chen:2021azt}, justifying the treatment of top-quark mass in the rest of this study. More details on this can be found in App.~\ref{sec:heftvssm}.

\subsection{POWHEG BOX}
\label{sec:powheg}
The \powhegbox~\cite{Frixione:2007vw,Alioli:2010xd,Jezo:2015aia} is an event generator
which produces Les Houches events~\cite{Boos:2001cv,Alioli:2013nda} with NLO accuracy that can be matched to parton showers according to the POWHEG method~\cite{Nason:2004rx}.
In \powhegbox, the following main simulation modes for ggF are available and used in this study:
\begin{itemize}
\item $H$ boson production at NNLO QCD~\cite{Monni:2019whf} implemented in the \minnlops framework (first achieved by reweighting $H$ in association with one extra jet in the \minlo framework in \cite{Hamilton:2013fea}). In this setup, the Higgs boson production without additional jets ($H+0j$) is NNLO accurate, $H+1j$ is NLO accurate and $H+2j$ is LO accurate.
\item $Hjj$ at NLO QCD~\cite{Campbell:2012am} implemented in the plain \powhegbox, with $H+0j$ and $H+1j$ not provided and with $H+2j$ at NLO precision.
\end{itemize}
We emphasize that the \minnlops-based prediction for the 2-jet final state is leading-order accurate.
Next-to-leading order precision is achieved by using the $Hjj$ simulation, which requires generation cuts or
Born suppression (see Sec.~\ref{subsec:matchingVariations} for details). While Born suppression works well at the
parton level, full event generation can fail if the simulation of multiple parton interactions is allowed to generate jets at transverse momenta larger than that of the hard process. Events with low transverse momentum jets in the hard process, which have large (and in principle unbounded) weights, can then be elevated to the 2-jet final state, which will result in distributions with seemingly random statistical outlier events.
We therefore decide to define the upper scale for both parton-shower and MPI evolution as the value given by the
\texttt{SCALUP} parameter. This technique is not available in \herwig, therefore comparisons including
\powhegbox+~\herwig will be limited to the parton level.

Note that a sample with all three multiplicities ($H+0j$, $H+1j$ and $H+2j$) provided at NLO accuracy can be obtained with an extension of the MiNLO method described in Ref.~\cite{Frederix:2015fyz}, but the corresponding code is currently not publicly available. 

\subsection{Pythia}
\label{sec:pythia}
\pythia~\cite{Sjostrand:2006za,Bierlich:2022pfr} is a multi-purpose particle-level event generator with a historically strong focus on soft physics.`
The non-perturbative evolution from colored partons to color-singlet hadrons is modeled using its distinctive feature, the string-fragmentation model \cite{Andersson:1983jt,Sjostrand:1984ic}.
In the present version 8.3, \pythia\ implements the \vincia\ antenna shower \cite{Brooks:2020upa} in addition to the default $p_\perp$-ordered ``simple shower'' \cite{Sjostrand:2004ef}\footnote{In versions prior to 8.316, also the \dire\ dipole-like parton shower \cite{Hoche:2015sya} was available.}.
For the simple shower, a dipole-recoil option is available, which replaces the separate incoherent initial-final and final-initial parton evolution by a common, coherent initial-final evolution \cite{Cabouat:2017rzi}{, in close correspondence to \vincia}.
{The \textsc{Apollo} parton shower \cite{Preuss:2024vyu} is not yet available for initial-state radiation. 
We therefore limit ourselves to the simple shower with and without using the dipole-recoil option.}
For each of the parton-shower algorithms, multi-parton interactions are fully interleaved with the shower evolution \cite{Sjostrand:2004ef}.
In this study, we use \pythia 8.315.

\pythia~8.3 implements a series of processes at LO.
Both the VBF $H+2j$ Higgs-production process and the gluon-fusion Higgs production process in the heavy-top limit are available at LO accuracy. 
In case of the latter, the simple shower algorithm also incorporates LO matrix-element corrections for the first emission, corresponding to the $H+1j$ configuration.
For matching and merging, \pythia relies on Les Houches events~\cite{Boos:2001cv,Alioli:2013nda} from external programs such as \textsc{MadGraph5}\_\textsc{aMC@NLO}~\cite{Alwall:2014hca}, \powhegbox, or \sherpa.
Multi-jet merging is implemented in the CKKW-L approach \cite{Lonnblad:2001iq,Lonnblad:2011xx}.
For NLO matching with the \powhegbox, dedicated \texttt{PowhegHooks} are available for all internal parton-shower algorithms to account for the mismatch between the evolution variable in \powhegbox and the shower evolution variable in \pythia \cite{Corke:2010zj}.
NLO matching with \textsc{MadGraph5}\_\textsc{aMC@NLO} is only available with the default simple shower with (non-default) global longitudinal recoil in final-state branchings. This was found to be problematic for certain simulations{~\cite{Cabouat:2017rzi,Ballestrero:2018anz,Jager:2020hkz,Hoche:2021mkv}}
and will therefore not be considered in this study.

\subsection{Herwig}
\herwig~\cite{Marchesini:1983bm,Marchesini:1987cf,Bahr:2008pv} is a general-purpose event generator with a historically strong focus on perturbative QCD and its connection to analytic resummation. Its distinctive feature is an angular-ordered parton shower (see Refs.~\cite{Gieseke:2003rz,Bewick:2019rbu} for a modern reformulation of such a shower) and a cluster hadronization model.
Its latest version, \herwig~7.3~\cite{Bellm:2019zci,Bewick:2023tfi}, also implements a dipole shower~\cite{Platzer:2009jq}, as well as an electroweak parton shower~\cite{Masouminia:2021kne}.
\herwig~7.3 allows for flexible NLO matching and merging capabilities. 
Thanks to its  \textsc{Matchbox} module~\cite{Platzer:2011bc}, both \textsc{Powheg}~\cite{Nason:2004rx} and \textsc{MC@NLO}~\cite{Frixione:2002ik} matching are offered, as well as unitarized merging~\cite{Bellm:2017ktr} (just for the dipole shower).
It also supports showering of Les Houches events generated either by the \powhegbox{} or by \textsc{MadGraph5}\_\textsc{aMC@NLO} (matching to \textsc{MadGraph5}\_\textsc{aMC@NLO} is supported only for the angular-ordered shower).

In this article, we  use the angular-ordered shower to process Les Houches events generated by the \powhegbox{}. In particular, we use the \textsc{powhegHerwig} plugin developed in Ref.~\cite{FerrarioRavasio:2023jck}, which is largely based on the Les Houches interface provided by the \textsc{ThePEG}~\cite{Lonnblad:2009zz}.

\subsection{Sherpa}
\sherpa~\cite{Gleisberg:2003xi,Gleisberg:2008ta,Bothmann:2019yzt,Sherpa:2024mfk} is a general-purpose
event generator with a strong focus on perturbative physics, in particular multi-jet merging methods. 
We use version 3.0 of the program for this study.
The matching of fixed-order NLO corrections to the parton shower is performed in the S-MC@NLO
approach~\cite{Hoeche:2011fd,Hoeche:2012ft}, while NLO computations are performed using 
\textsc{Amegic}~\cite{Krauss:2001iv,Gleisberg:2007md}, \textsc{Comix}~\cite{Gleisberg:2008fv}, and 
MCFM~\cite{Campbell:1999ah,Campbell:2011bn,Campbell:2015qma,Campbell:2019dru,Campbell:2021vlt}.
The virtual corrections have been cross checked against \textsc{OpenLoops}~\cite{Cascioli:2011va,Buccioni:2017yxi,Buccioni:2019sur}.
We use a modified version of the Catani-Seymour dipole shower algorithm (CSS)~\cite{Schumann:2007mg},
which was cross-checked against the \textsc{Dire} parton shower~\cite{Hoche:2015sya} (for S-MC@NLO only).
The \textsc{Dire} shower has been deprecated in \sherpa, while the NLO matching~\cite{Hoche:2025gsb} for the
new \textsc{Alaric} parton shower~\cite{Herren:2022jej,Assi:2023rbu,Hoche:2024dee} is not yet available
for initial-state evolution. Therefore we only present predictions using the default CSS algorithm.
The NLO multi-jet merging is based on~\cite{Gehrmann:2012yg,Hoeche:2012yf}.
Hadronization in \sherpa is provided by a cluster fragmentation model~\cite{Winter:2003tt,Chahal:2022rid},
and multiple parton scattering as a model of the underlying event is simulated according
to~\cite{Alekhin:2005dx,Alekhin:2005dy}.

\section{Comparison to typical LHC simulations}
\label{sec:previous_simulations}
Before performing a detailed analysis of theoretical uncertainties, we compare our new predictions
to those obtained from a setup routinely employed by the experimental collaborations.
This provides the context for the discussion of the true uncertainties in Sec.~\ref{sec:uncertainties}.
The event generator conventionally used by ATLAS and CMS for the
ggF Higgs boson production background to VBF Higgs boson production was NNLOPS~\cite{Hamilton:2013fea,Hamilton:2015nsa}. It has now
largely been replaced by the more modern \minnlops generator~\cite{Monni:2019whf,Niggetiedt:2024nmp},
which gives formally equivalent results while being substantially more efficient, due to the 
elimination of the reweighting step. Both NNLOPS and \minnlops are usually interfaced with
the \herwig and \pythia Monte-Carlo programs. The estimated uncertainty in the
ggF Higgs boson background is dominated by the observed differences between the predictions of
NNLOPS+\pythia and NNLOPS+\herwig, which arise from both perturbative and non-perturbative QCD effects
{as well as technical differences}.

A key point is that these are the differences observed by using the simulations ``out of the box''
(albeit with tuned non-perturbative parameters from each experiment),
and not differences resulting from true discrepancies in the physics models of the event generators.
{One prominent example of this is the inconsistent usage of the CMW scale scheme~\cite{Catani:1990rr},
which incorporates the resummed single-logarithmic corrections due to the two-loop cusp anomalous 
dimension~\cite{Kodaira:1982az} in the parton shower. In \pythia, this correction is optional and
it is not included by default, while in \herwig and \sherpa it is enabled by default. The missing
single-logarithmic terms in \pythia can be compensated for by using a larger value of the strong
coupling, which is often misinterpreted as \pythia actually requiring a different coupling definition.}
In this section, we will use the ATLAS and CMS setups of \minnlops for comparison to the ones we
recommend as a new baseline to reach the highest theoretical precision in $H+2$ jets from ggF.

\subsection{Observable definitions}
\label{sec:observables}
We will discuss the following four observables, which are sensitive to the characteristics
of the Higgs-plus-two-jet final state:
\begin{itemize}
    \item the transverse momentum of the Higgs boson, $p_{\mathrm{T},\text{H}}$;
    \item the invariant mass of the two leading jets, $m_{jj}$;
    \item the rapidity separation of the two leading jets, $\Delta y_{jj}$;
    \item the azimuthal separation of the two leading jets, $\Delta \phi_{jj}$.
\end{itemize}
Additional data are collected on \zenodoplotref, together with the
Rivet~\cite{Buckley:2010ar,Bierlich:2019rhm,Bierlich:2024vqo} analysis routine.
{A selection of further figures not contained in the main text below is given in Appendix~\ref{sec:additionalFigures}.}

The jets are clustered using the \fastjet implementation of the anti-$k_T$
algorithm~\cite{Cacciari:2008gp}, and must satisfy the transverse momentum cut
$p_T^{\text{jet}} > 30\ \text{GeV}$.\footnote{
We have verified that the conclusions of our study are unchanged when working with
more exclusive phase-space cuts, in particular a Higgs boson rapidity cut of
$|y_{\text{H}}|<2.4$ and a jet rapidity cut of $|y^{\text{jet}}|<4.4$.}
We primarily consider jets with a radius $R$ of 0.4, but also provide results for radii
of 0.3, 0.6 and 0.8, to quantify the impact on selected differential observables.\footnote{%
  Our study of the VBF Higgs boson production process in Ref.~\cite{Buckley:2021gfw}
  examined jet $R$-values from 0.3 to 1.0,
  and indicated a strong $R$-dependence due to the kinematic requirements on the two jets.
  The R-dependence increased upon the imposition of the VBF phase space cuts, so it is
  important to check if the same effects occur for the background. Note that the two leading jets in 
  VBF Higgs boson production are primarily quark jets, while in ggF they are primarily gluon jets.
  This may led to some differences in the $R$-dependence. }
Jets are ranked by the magnitude of their transverse momentum.
{As a diagnostic tool, we sometimes also investigate observables in the Higgs+1~jet selection,
which is defined as follows: Jets are clustered
using the \fastjet implementation of the anti-$k_T$ algorithm~\cite{Cacciari:2011ma},
and must satisfy $p_T^{\text{jet}} > 30\ \text{GeV}$.}

To distinguish between normalization and shape differences of the predictions, 
all results will be normalized to the inclusive cross section for the $H+2j$ final state,
defined at the particle level{, see Appendix~\ref{sec:additionalFigures}}. This means in particular that a normalization difference
can occur through the different behavior of the parton shower simulation, for example
if the parton shower shifts a jet above/below the $p_\mathrm{T}$ cut by means of additional radiation
and/or recoil effects. The cross section is inferred from a relatively inclusive distribution,
which we choose to be the Higgs boson rapidity distribution in the two jet final state.

\subsection{Results of initial comparison}
\label{sec:minnlo_comparison}
We are now ready to compare results from the \minnlops simulation as used by ATLAS and CMS
to those from the tools introduced in Sec.~\ref{sec:setup}. 
The \minnlops calculation is matched to the default \pythia\ parton shower (``simple shower''),
both with and without the ``dipole-recoil'' option, and to the \herwig angular-ordered parton shower.
The results are compared to a MEPS@NLO simulation from \sherpa, including the $H+0j$, $H+1j$, and $H+2j$
final states at NLO precision, and up to $H+5j$ at LO accuracy. This sample will be dubbed MEPS@NLO
in the following. All predictions are at parton level, without multi-parton interactions
and non-perturbative corrections. We also use a fixed-order $H+1j$ NNLO calculation 
from \nnlojet as a reference.

\begin{figure}[t]
    \centering
    \includegraphics[width=0.425\linewidth]{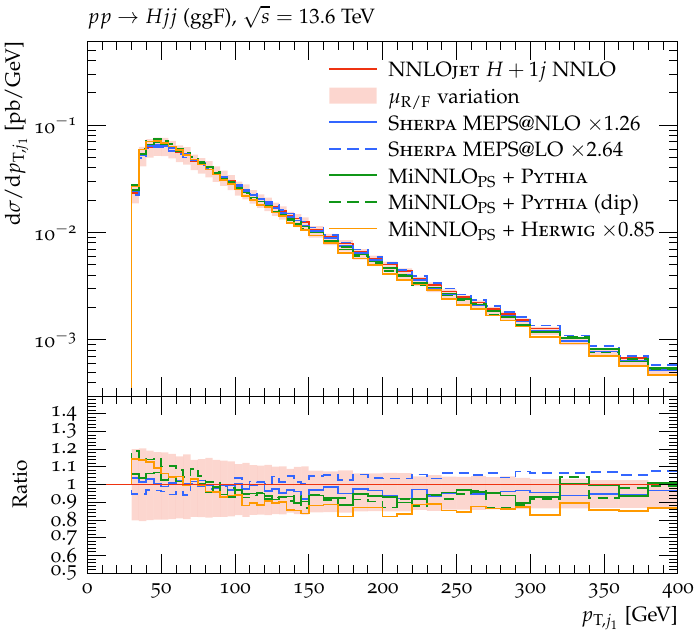}
    \includegraphics[width=0.425\linewidth]{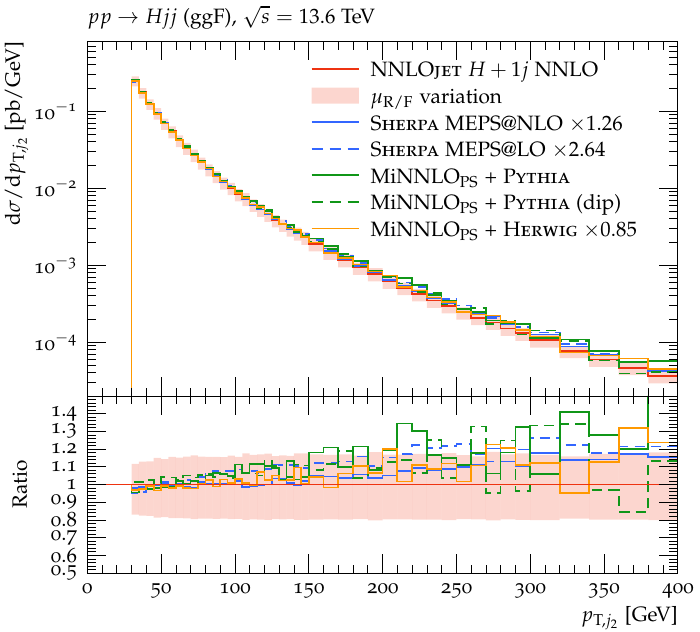}\\
    \caption{Comparison of the inclusive event-generation setups MEPS@NLO and \minnlops
    to the \nnlojet $H+1j$ NNLO prediction in the fiducial $H+2j$ phase space for the
    transverse momentum of the leading (left) and sub-leading (right) jet.
    Scale factors are derived from the \nnlojet Higgs rapidity distribution,
    computed at NLO precision.
    The 2-jet final state is predicted at NLO-accuracy by MEPS@NLO and NNLOJET,
    and at LO accuracy in \minnlops.}
    \label{fig:jetSpectra}
\end{figure}
{
We find approximate agreement of the rate in the Higgs rapidity distribution of the $H+2j$ selection
(cf.\ Fig.~\ref{fig:higgsRapidity}) between \nnlojet and \minnlops+~\pythia, whereas the \sherpa MEPS@NLO sample
requires a scale factor of $1.26$, and the \minnlops+~\herwig sample a scale factor of $0.86$.
In the case of the \sherpa prediction, the scale factor is consistent with the NNLO to NLO
ratio of the inclusive Higgs boson production cross section~\cite{Anastasiou:2016cez},
which is needed to bring the inclusive jet rates in the MEPS@NLO result into approximate agreement
with the prediction from \nnlojet. We will make use of these scale factors in most of the following
comparisons.

In Fig.~\ref{fig:jetSpectra} we show the leading and sub-leading jet transverse momentum
spectra in the inclusive $H+2j$ selection. The reference prediction stems from NNLOjet,
which provides NLO precision for this observable. The \minnlops simulation models it
at LO precision. The MEPS@NLO simulation predicts it at NLO precision. To gauge the difference
between LO and NLO accurate simulations, we also show a leading-order multi-jet merged result, labeled MEPS@LO.
The MEPS@NLO, MEPS@LO and \minnlops results all include Sudakov suppression effects, and both
MEPS@NLO and MEPS@LO include higher-order tree-level corrections up to $H+5$jets.
Figure~\ref{fig:jetSpectra} shows that the shape of the scaled MEPS@NLO and \minnlops+\pythia
results agree with the fixed-order calculation of the leading jet transverse momentum within $\pm 5\%$
over the entire fiducial phase space, and in the medium hard region $p_{T,j1}\lesssim m_H$ in particular.
Deviations between MEPS@NLO and MEPS@LO are of similar size at transverse momenta below $m_H$,
but they can reach about 10\% in the hard region $p_T\gtrsim m_H$. The shape of the \minnlops+\pythia~(dip)
and \minnlops+\herwig predictions differs significantly from the fixed-order result and the
other two predictions. The situation is different for the sub-leading jet transverse momentum,
where the MEPS@NLO and \minnlops+\herwig prediction agree with the fixed-order result from
NNLOjet within five percent up to $\sim 250$~GeV, and within about 10\% up to 400~GeV.
Here, the \minnlops+\pythia and \minnlops+\pythia~(dip) predictions show deviations up to 20\%
already in the region below $\sim 250$~GeV. We will discuss similar features in more detail
in Sec.~\ref{subsec:inc_vs_nlops}, and in particular when comparing the distributions
in Fig.~\ref{fig:incVsNLOPSphijj}.}

\begin{figure}[t]
    \centering
    \includegraphics[width=0.4\linewidth]{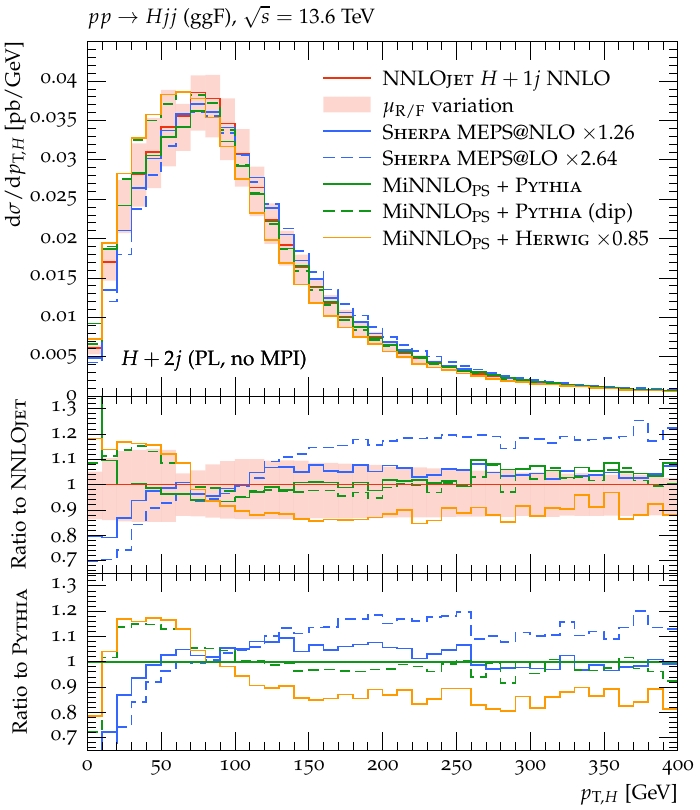}
    \includegraphics[width=0.4\linewidth]{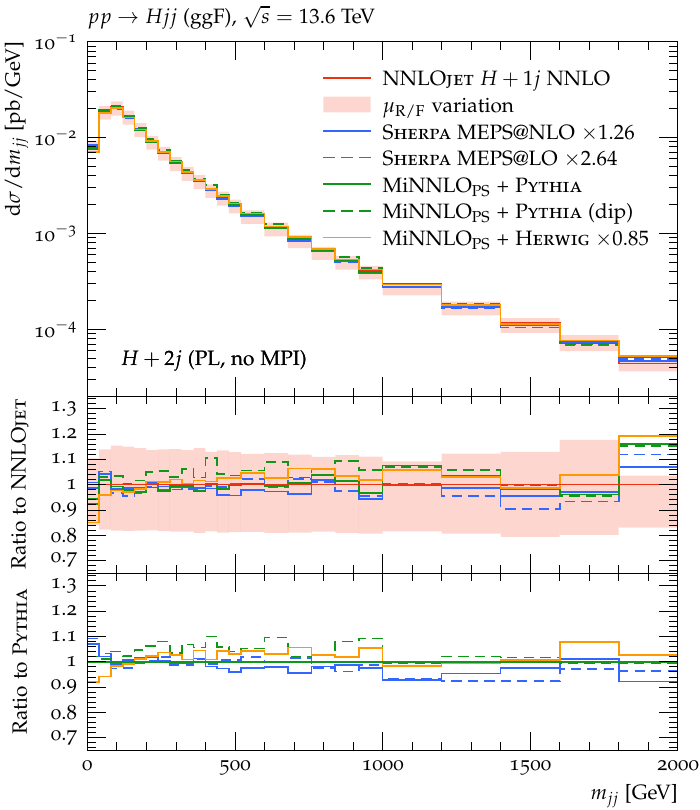}
    \includegraphics[width=0.4\linewidth]{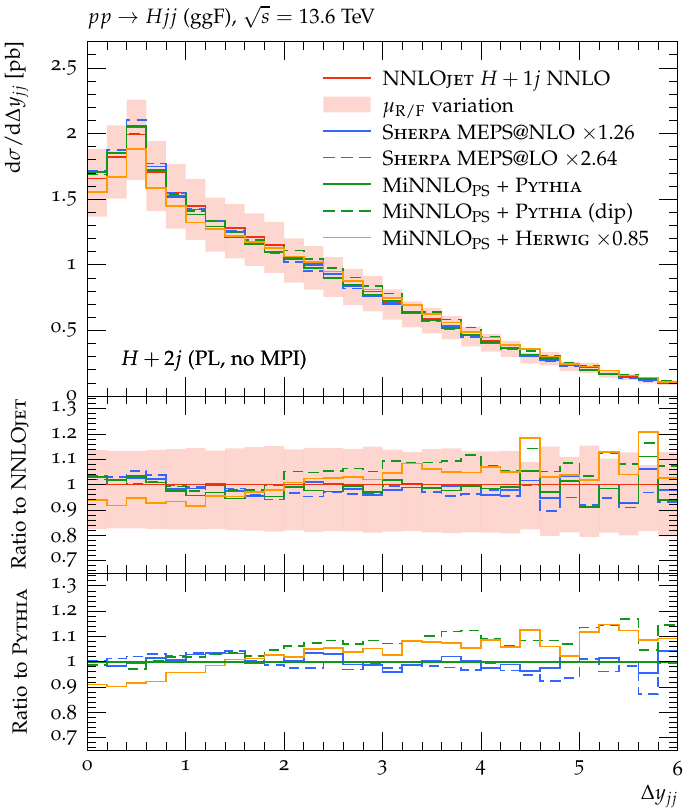}
    \includegraphics[width=0.4\linewidth]{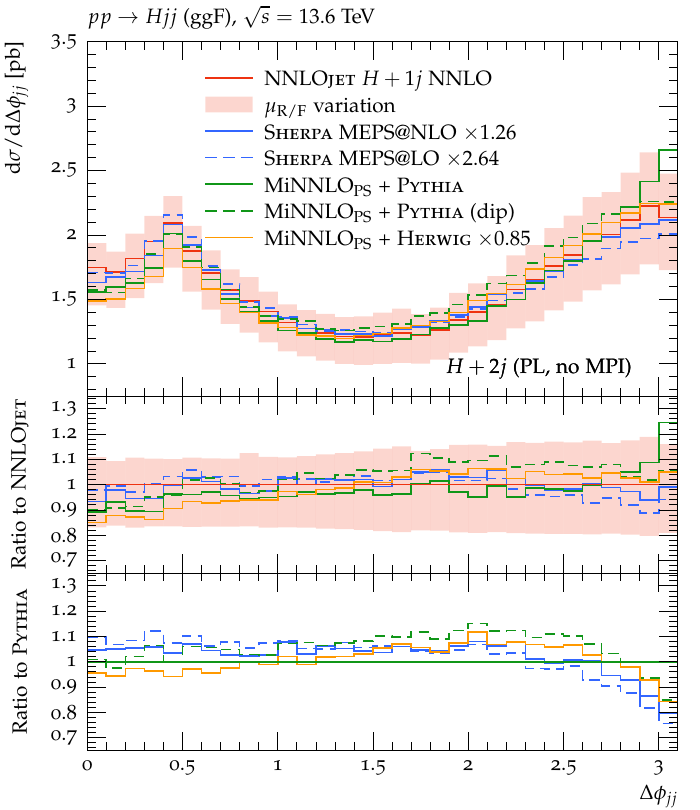}
    \caption{Comparison of the inclusive event-generation setups MEPS@NLO and \minnlops
    to the \nnlojet $H+1j$ NNLO prediction in the fiducial $H+2j$ phase space for the
    Higgs transverse momentum $p_{\mathrm{T},H}$ (top left), the dijet invariant mass
    (top right), the dijet rapidity separation $\Delta y_{jj}$ (bottom left), and the
    dijet azimuthal angle separation $\Delta\phi_{jj}$ (bottom right).
    Scale factors are derived from the \nnlojet Higgs rapidity distribution,
    computed at NLO precision.
    The 2-jet final state is predicted at NLO-accuracy by MEPS@NLO and NNLOJET,
    and at LO accuracy in \minnlops.
    }
    \label{fig:inclusive}
\end{figure}
Figure~\ref{fig:inclusive} shows {the comparison of the \nnlojet reference to the MEPS@NLO and \minnlops predictions, which will be referred to as inclusive event-generation setups in the following}. The middle panels of each sub-figure
display the ratio to the inclusive fixed-order calculation from \nnlojet, while the bottom panels
compare the event-generator predictions, normalized to the {result from \minnlops+~\pythia with default recoil}.
The rapidity separation $\Delta y_{jj}$ and the invariant mass distribution $m_{jj}$ of the two leading
jets, shown in Figure~\ref{fig:inclusive}, agree well between the \sherpa, \pythia, and \herwig setups. 
All results also agree well with \nnlojet.
The distributions of the azimuthal separation $\Delta \phi_{jj}$ generally show good agreement
between the different predictions in the region of small and intermediate separations, where they lie
within a 10\% envelope of each other. At large jet separations, $\Delta\phi_{jj} \gtrsim 2.5$,
the \minnlops+~\pythia prediction is larger than all other predictions by about 30\%, including
the \minnlops+~\pythia results using the dipole-recoil option. We will discuss this in more
detail in Fig.~\ref{fig:incVsNLOPSphijj}. The Higgs transverse-momentum
distribution $p_{\mathrm{T},\mathrm{H}}$ shows the largest shape differences between the different simulations.
At low transverse momentum, $p_{\mathrm{T},\mathrm{H}} \lesssim 50~\gev$, both \minnlops+~\pythia results
and \minnlops+~\herwig exceed the \nnlojet result by about 20\%, with close agreement between
the \minnlops predictions obtained with \herwig and with the \pythia dipole-recoil shower.
The \sherpa MEPS@NLO prediction is about 20\% lower than the \nnlojet result.
Starting from $p_{\mathrm{T},\mathrm{H}} \gtrsim 100~\gev$, both \minnlops+~\pythia predictions agree
very well with \nnlojet, whereas the \sherpa MEPS@NLO and \minnlops+~\herwig predictions create
an envelope around the fixed-order result, with \sherpa MEPS@NLO being larger by about 10\% and
\minnlops+~\herwig being smaller by about the same amount. As can be inferred from the middle panes
in all four figures, the event-generator predictions broadly lie within the seven-point scale variation
uncertainty band of the fixed-order calculation, with some exceptions towards larger values
of the observables.

Considering the observations in Figs.~\ref{fig:jetSpectra} and~\ref{fig:inclusive},
one may be tempted to conclude that the modeling of the irreducible background
to VBF Higgs production suffers from large theory uncertainties.
This is in fact not the case. The event-generator predictions compared so far have rather different
formal accuracy in the fiducial $H+2j$ phase space, and therefore provide a different level of theoretical
control over the observables of interest. 
{In particular, the \minnlops and MEPS@LO results are formally leading order accurate for these
observables. However, the \minnlops results differ between \herwig and \pythia for all observables
in a way that is not suggested by the difference between the MEPS@LO and the MEPS@NLO results.
We are therefore led to the conclusion that the difference in the simulations is not simply
due to a difference in fixed-order higher-order corrections. We will focus on a practical solution
to mitigate such problems with modeling the $H+2j$ final state, which also relies on the simulations
with the highest parametric precision for the $H+2$jet final state.}

\section{Assessment of theory uncertainties}
\label{sec:uncertainties}
In the following, we will use a consistent baseline for the hard, perturbative calculation
-- $H+2j$ at NLO precision -- and assess the influence of different theory variations on the size
of the theoretical uncertainties in the fiducial $H+2j$ phase space. Throughout the discussion,
we will focus on the same four observables as in Fig.~\ref{fig:inclusive}, with additional results
available at \zenodoplotref. In Sec.~\ref{subsec:showerVariations}, we compare parton-shower variations
in $H+2j$ NLO-matched distributions at the parton level. Sec.~\ref{subsec:matchingVariations} focuses
on matching variations in \powhegbox, while the influence of dedicated \pythia tunes and different hadronization models in \sherpa is discussed
in Sec.~\ref{subsec:tuneVariations}. A study of the impact of the Higgs effective
theory approximation can be found in App.~\ref{sec:heftvssm}  and  a study of the jet-radius dependence can be found in App.~\ref{sec:coneSizeChoice}.

\subsection{Parton-shower variations}
\label{subsec:showerVariations}
In this section, we compare the influence of different parton-shower algorithms,
matched to a $H+2j$ NLO calculation. We consider a total of four different showers
in \sherpa, \pythia, and \herwig. In \sherpa, we use the Catani-Seymour shower
matched to the NLO $H+2j$ calculation using the S-MC@NLO scheme;
in \pythia, we employ the default $p_\mathrm{T}$-ordered ``simple'' shower
as well as the simple shower with dipole recoil, both matched to the NLO $H+2j$
calculation using \powhegbox; in \herwig, we use the angular-ordered shower
matched to the NLO $H+2j$ calculation in the POWHEG scheme using \powhegbox.
A discussion of uncertainties pertaining to the matching scheme is given
in Sec.~\ref{subsec:matchingVariations}.
In all cases, we consider parton-level predictions without modeling the
underlying event and we do not include hadronization effects.
The impact of these corrections is addressed in
Sec.~\ref{subsec:tuneVariations}.

\begin{figure}[t]
    \centering
    \includegraphics[width=0.425\linewidth]{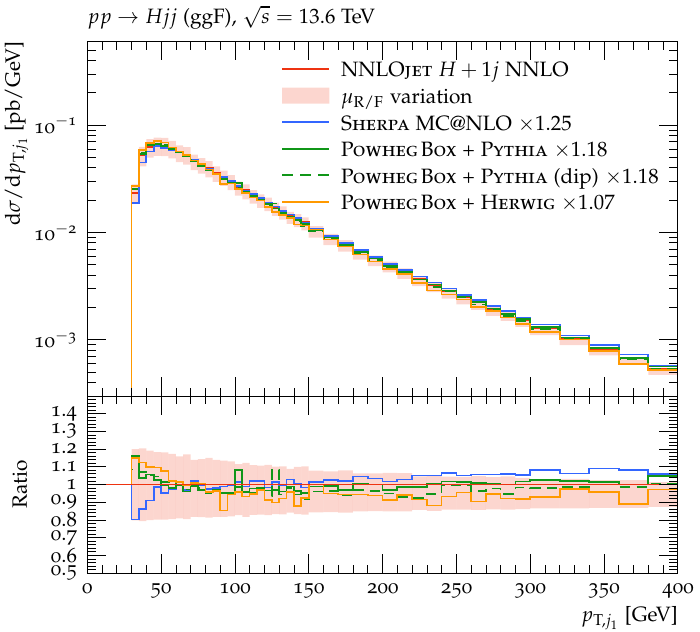}
    \includegraphics[width=0.425\linewidth]{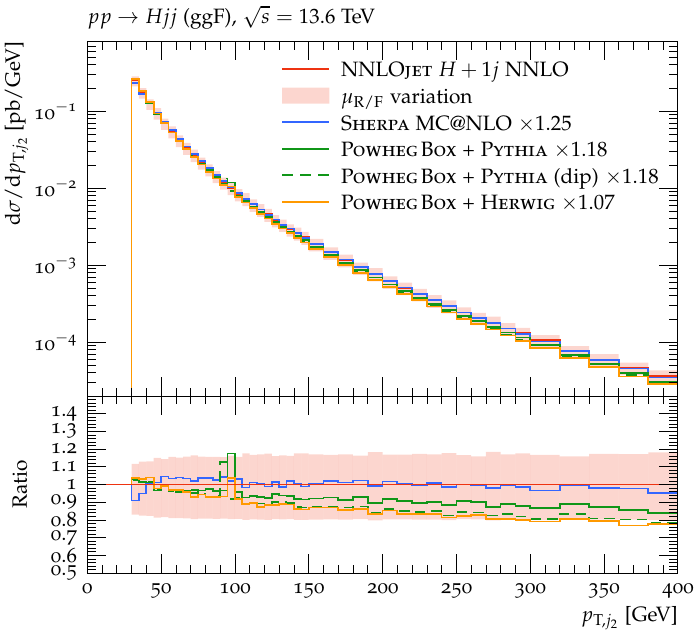}
    \caption{Comparison of the \sherpa MC@NLO, \powhegbox + \pythia, and \powhegbox + \herwig $H+2j$ NLO+PS predictions to the \nnlojet $H+1j$ NNLO prediction in the fiducial $H+2j$ phase space for the transverse momentum of the leading (left) and sub-leading (right) jet.
    Scaling factors are derived from the \nnlojet Higgs rapidity distribution, computed at NLO precision. All predictions are NLO-accurate for the two jet final state. No non-perturbative corrections have been applied.}
    \label{fig:jetSpectra2}
\end{figure}

\begin{figure}[t]
    \centering
    \includegraphics[width=0.4\linewidth]{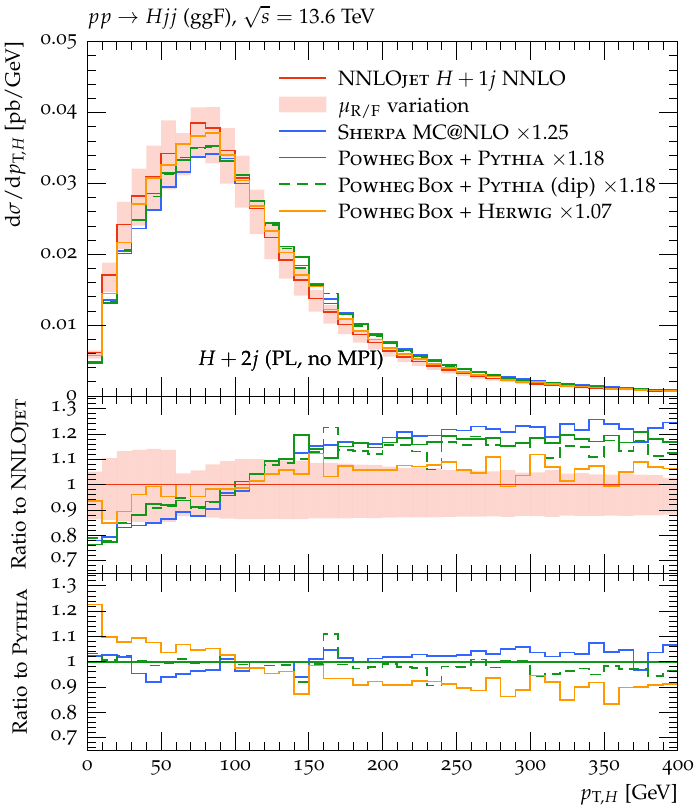}
    \includegraphics[width=0.4\linewidth]{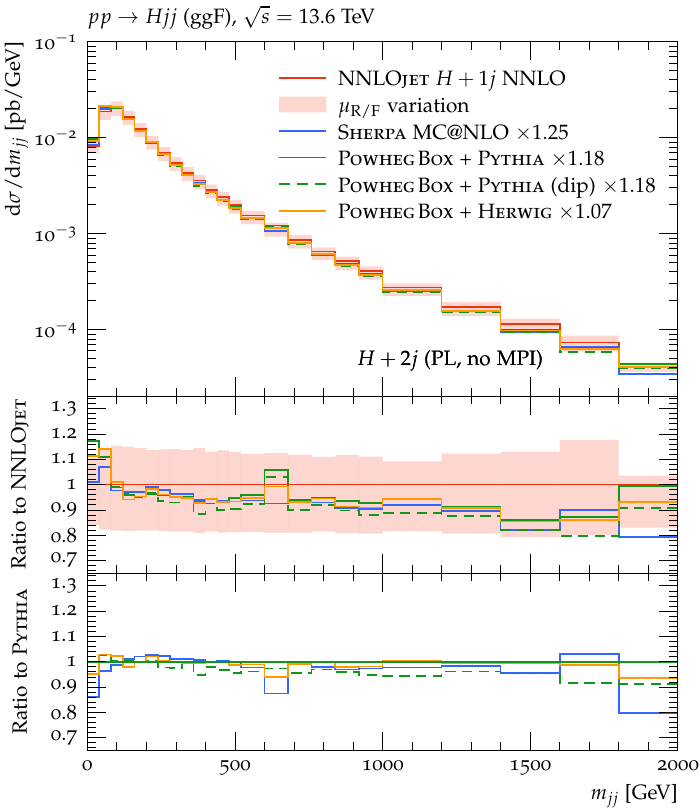}
    \includegraphics[width=0.4\linewidth]{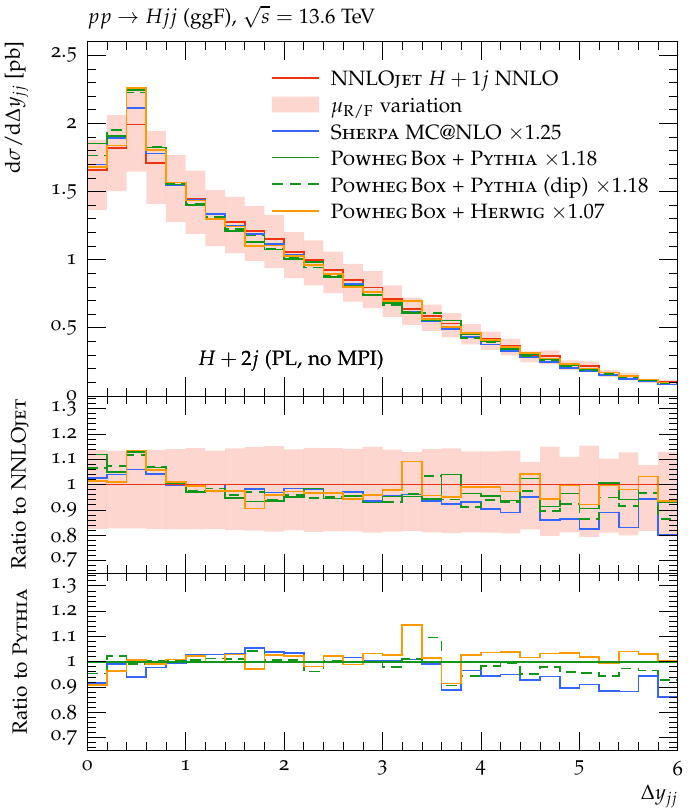}
    \includegraphics[width=0.4\linewidth]{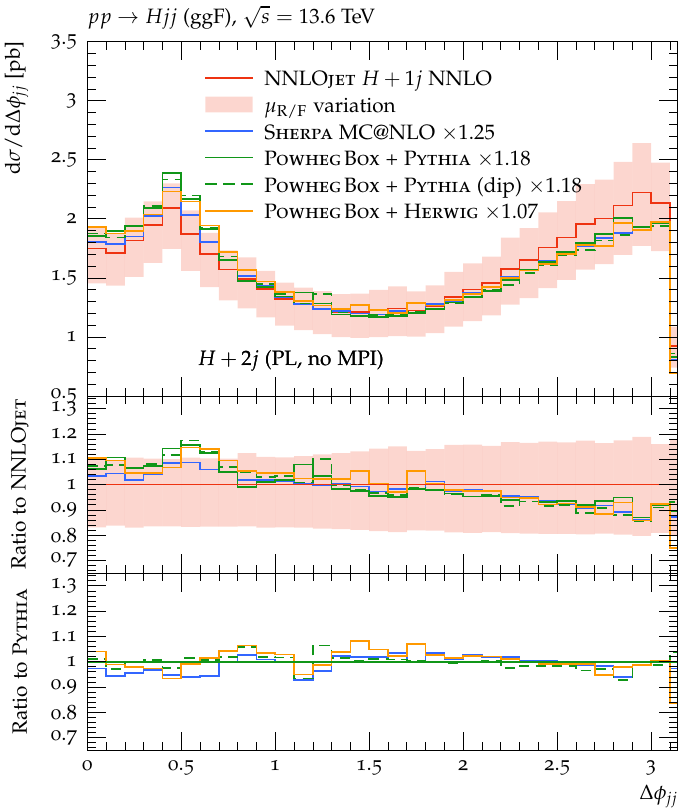}
    \caption{Comparison of the \sherpa MC@NLO, \powhegbox + \pythia, and \powhegbox + \herwig $H+2j$ NLO+PS predictions to the \nnlojet $H+1j$ NNLO prediction in the fiducial $H+2j$ phase space for the Higgs transverse momentum $p_{\mathrm{T},H}$ (top left), the dijet invariant mass (top right), the dijet rapidity separation $\Delta y_{jj}$ (bottom left), and the dijet azimuthal angle separation $\Delta\phi_{jj}$ (bottom right).
    Scaling factors are derived from the \nnlojet Higgs rapidity distribution, computed at NLO precision. All predictions are NLO-accurate for the two jet final state. No non-perturbative corrections have been applied. }
    \label{fig:showerVariations}
\end{figure}

As before, we scale all distributions to the \nnlojet rate based on the
Higgs rapidity spectrum, yielding scale factors of $1.26$ in the case
of \sherpa, $1.18$ for both \pythia showers, and $1.06$ for \herwig.
The resulting parton-shower variations are shown in Fig.~\ref{fig:showerVariations},
including the \nnlojet $H+1j$ NNLO prediction as a baseline. Besides the main
distributions, we also show two ratio panels, with the first (middle panel)
comparing the matched predictions to the \nnlojet fixed-order reference,
and the second (bottom panel) showing the difference between the matched
parton-shower predictions, normalized to the \powhegbox+~\pythia result.
All predictions in Figs.~\ref{fig:jetSpectra2} and~\ref{fig:showerVariations}
acquire formal NLO accuracy in $H+2j$ configurations. 

Figure~\ref{fig:jetSpectra2} shows similar agreement between the NLO matched
results as between MEPS@NLO and \minnlops in Fig.~\ref{fig:jetSpectra}.
As can be seen from Fig.~\ref{fig:showerVariations}, we find percent-level
agreement between the four matched parton-shower algorithms for the
$\Delta y_{jj}$, $\Delta \phi_{jj}$, and $m_{jj}$ distributions.
For the $p_{\mathrm{T},\mathrm{H}}$ distribution, both \pythia showers agree very well
with \sherpa, while the \herwig angular-ordered shower differs from these
at large transverse momentum, $p_{\mathrm{T},\text{H}} > 100~\gev$.
{Except for the Higgs transverse-momentum distribution}, the matched parton-shower results are within the scale uncertainties
of the fixed-order calculation, although the agreement with \nnlojet is considerably
better for $\Delta y_{jj}$, $m_{jj}$ and $\Delta\phi_{jj}$ than for
$p_{\mathrm{T},\mathrm{H}}$, with the latter showing the largest differences
of up to 20\%.

\subsection{Matching-scheme variations}
\label{subsec:matchingVariations}
{We now turn to }investigate ambiguities
in the matching scheme.  Since different matched parton-shower predictions
generally agreed very well, we focus on \powhegbox+~\pythia, where we expect
the largest differences arising from ambiguities in the multiplicative
POWHEG approach and the interfacing to an external shower generator.
Specifically, we study the influence of the following \powhegbox
event-generation parameters:
\begin{itemize}
\item the cut-off scale of the first emission (handled by \powhegbox), \texttt{ptsqmin};
\item the scale choice in real contributions, \texttt{btlscale},
namely whether to use the real kinematics of the corresponding
Born projection for the computation of the renormalization
and factorization scale (see Sec.~3 of Ref.~\cite{FerrarioRavasio:2023jck});
\item \texttt{hdamp}, i.e.\ the transverse-momentum scale used to
separate regular real events from soft ones
(see Sec.~5 of Ref.~\cite{Alioli:2010xd} for further details).
\end{itemize}
In addition, there is some ambiguity in how the scale mismatch between
the first emission in \powhegbox and \pythia is treated \cite{Corke:2009tk}. 
Natively, \pythia uses \texttt{PowhegHooks} to veto emissions harder than
the first emission of \powhegbox. By default, however, the $H+2j$ process
in \powhegbox regulates the $H+2j$ process by a procedure known as Born
suppression, in which infrared divergent configurations are suppressed.
This approach differs from the usual generation cuts in that events
with singular configurations are not entirely excluded from the event
sample but enter the sample with a reduced (or vanishing) weight.
Typically, these singular events are excluded from the analysis by the
fiducial cuts, which separate out the $H+2j$ topology.
However, we found that in both \pythia and \herwig this approach suffers
from deteriorating statistical precision, as jets created by multi-parton
interactions can migrate (previously suppressed) infrared-divergent events
into the fiducial phase space (see also Sec.~\ref{sec:powheg}).
By default, we have therefore decided against using \texttt{PowhegHooks}
in \pythia for our main study, meaning that both the
parton-shower and MPI evolution is started at the \powhegbox
\texttt{SCALUP} scale without further vetoes.
A similar approach does not seem viable in the underlying-event simulation
in \herwig, where the MPI starting scale cannot be altered.
This excludes full event simulations, including hadronization and MPI,
with \powhegbox+~\herwig from our study. We hope that it will be possible to implement this feature in \herwig in the future. 

In order to gauge the importance of vetoed showers in our predictions,
we have implemented generation cuts in the \powhegbox $H+2j$ process
and compare the following event-generation setups:
\begin{itemize}
\item default Born suppression without \texttt{PowhegHooks}:
the importance sampling is altered so to have many events
at large $p_\mathrm{T}$, and few events with large weight at small $p_\mathrm{T}$;
\item tighter generation cuts without \texttt{PowhegHooks};
\item tighter generation cuts with \texttt{PowhegHooks} and \texttt{POWHEG:pThard = 0};
\item tighter generation cuts with \texttt{PowhegHooks} and \texttt{POWHEG:pThard = 2}.
\end{itemize}
In this context, the \texttt{POWHEG:pThard} setting defines how the
veto scale is calculated in the \texttt{PowhegHooks}, with
\texttt{POWHEG:pThard = 0} corresponding to the scale of the
\powhegbox emission, \texttt{SCALUP}, and \texttt{POWHEG:pThard = 2}
amounts to recalculating the veto scale as the minimal $p_\mathrm{T}$
across the entire event. With properly adjusted parton-shower and
MPI starting scales, the first two options should fully agree.
The third option provides a proxy for the importance of vetoed showers
in our setup, while the last option corresponds to an alternative
veto-scale choice in \texttt{PowhegHooks}, which was recently adopted
in the ATLAS experiment to quantify the evolution-scale mismatch
between \powhegbox and \pythia \cite{ATLAS:2023kov}.
It should be noted that sizable differences between the latter two
options have been observed for the VBF process in \cite{Hoche:2021mkv}.

\begin{figure}[t]
    \centering
    \includegraphics[width=0.4\linewidth]{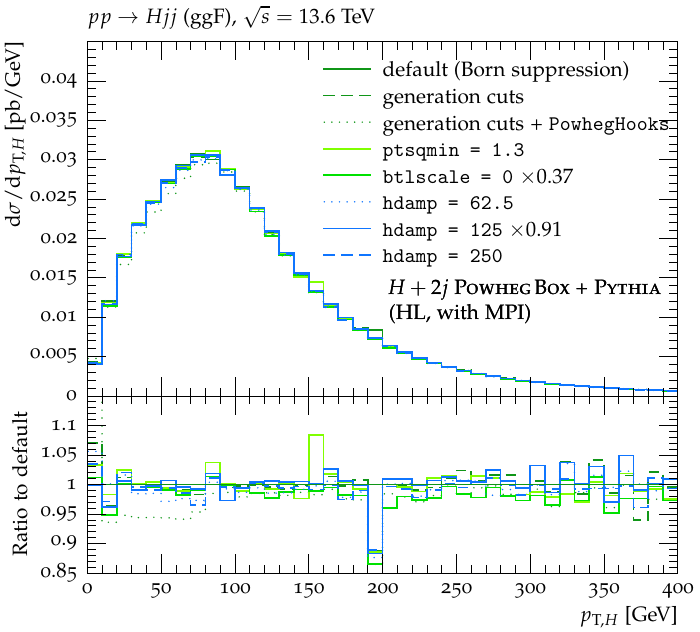}
    \includegraphics[width=0.4\linewidth]{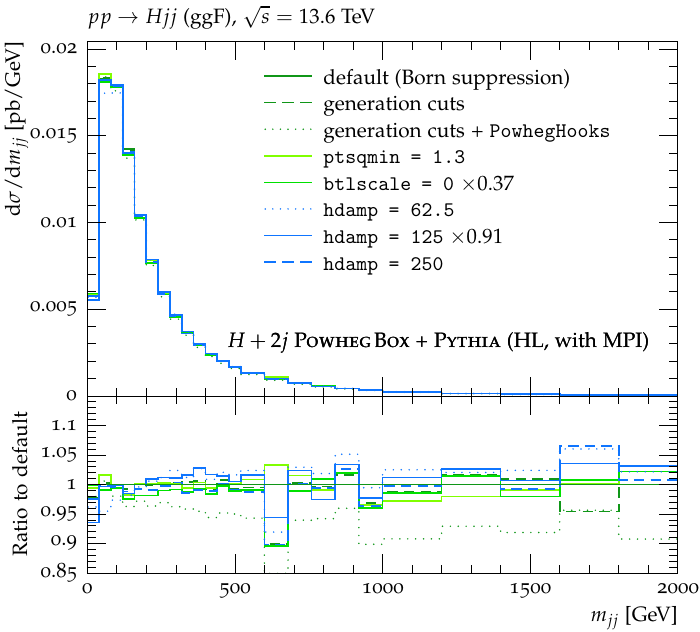}
    \includegraphics[width=0.4\linewidth]{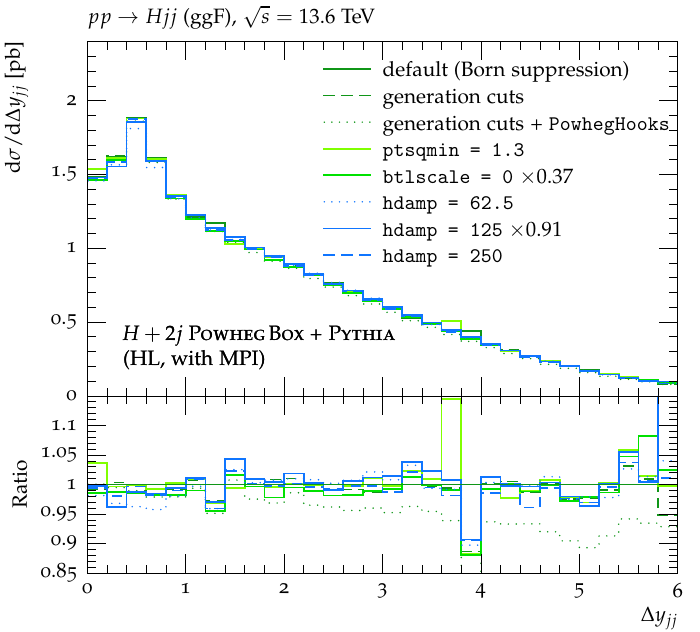}
    \includegraphics[width=0.4\linewidth]{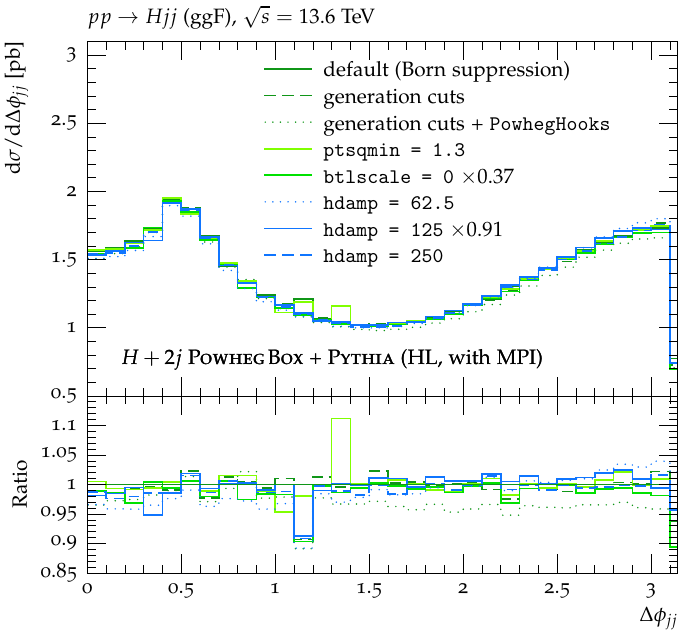}
    \caption{Impact of matching-scheme ambiguities in \powhegbox+~\pythia $H+2j$ NLO+PS matching in the fiducial $H+2j$ phase space for the Higgs transverse momentum $p_{\mathrm{T},\mathrm{H}}$ (top left), the dijet invariant mass (top right), the dijet rapidity separation $\Delta y_{jj}$ (bottom left), and the dijet azimuthal angle separation $\Delta\phi_{jj}$ (bottom right). 
    Scaling factors are derived from the default \powhegbox+~\pythia Higgs rapidity distribution.}
    \label{fig:matchingVariations}
\end{figure}
A comparison of the different matching options is shown in
Fig.~\ref{fig:matchingVariations}, where the lower panel compares
the various settings to the default approach using Born suppression
without \texttt{PowhegHooks}. In case of the \texttt{btlscale = 0}
(i.e.\ using the real kinematics to compute the scales entering the real cross section)
and \texttt{hdamp = 125}  predictions, we again apply scaling factors
of $0.37$ and $0.91$, respectively, based on the Higgs rapidity distribution.
We observe excellent agreement in the shape of all distributions,
with differences well below the percent-level.
The largest differences are found with the \texttt{hdamp = 62.5}
predictions, which differs from the other predictions by up to 10\%
in the tails of the $m_{jj}$, $\Delta y_{jj}$, and $\Delta \phi_{jj}$
distributions, and by about 5\% in the region of small Higgs
transverse momentum, $p_{\mathrm{T},H} \lesssim 75~\gev$.
This emphasizes that within the freedom of the \powhegbox
matching implementation, consistent results are obtained as long
as the parameters do not take on extreme values.
We find that the \texttt{btlscale} setting has a small impact
on the shape of distributions,
while a small value of \texttt{hdamp} can cause an important
shape difference at large $m_{jj}$/$\Delta y_{jj}$.
This is due to the fact that, in the POWHEG method,
for relatively large transverse-momentum radiation,
the predicted rate is effectively equivalent to the real
cross section multiplied by the ratio of the inclusive
NLO/LO cross section, which is positive for the process at hand.
When \texttt{hdamp} is used, radiation harder than this parameter
is treated separately and effectively does not include this factor.

\subsection{Tune variations}
\label{subsec:tuneVariations}
We next turn to study uncertainties arising from differences in the
modeling of non-perturbative and underlying-event effects. To this end,
we consider the $H+2j$ MC@NLO prediction obtained with \sherpa in
conjunction with either \sherpa's cluster-fragmentation~\cite{Chahal:2022rid}
or the string-fragmentation model in \pythia, interfaced at runtime.
In addition, we consider three different tunes in the \powhegbox+~\pythia
setup with the default shower (without dipole recoil): the default
Monash2013 tune in \pythia \cite{Skands:2014pea}, the ATLAS AZNLO tune
\cite{ATLAS:2014alx}, and the CMS CP5 tune~\cite{CMS:2019csb}.

\begin{figure}[t]
    \centering
    \includegraphics[width=0.4\linewidth]{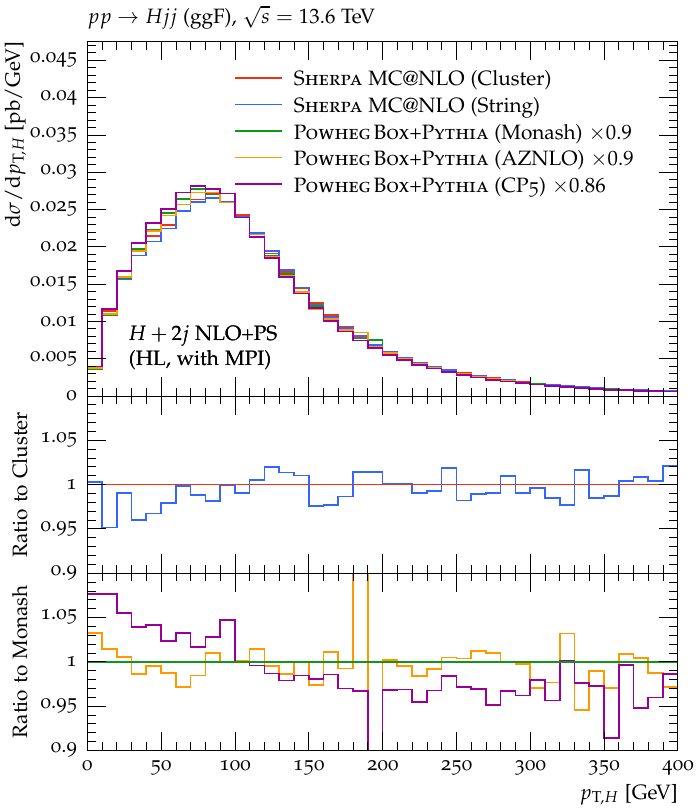}
    \includegraphics[width=0.4\linewidth]{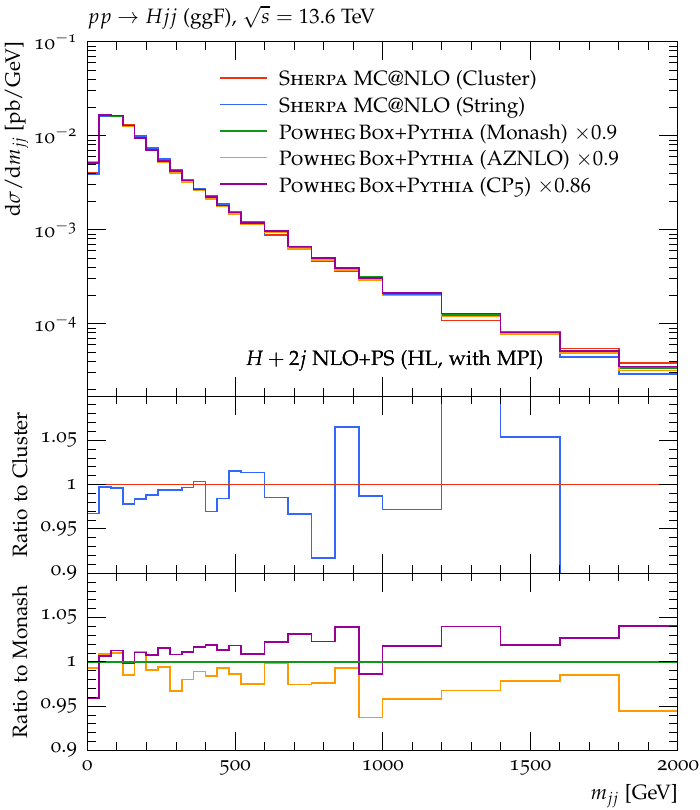}
    \includegraphics[width=0.4\linewidth]{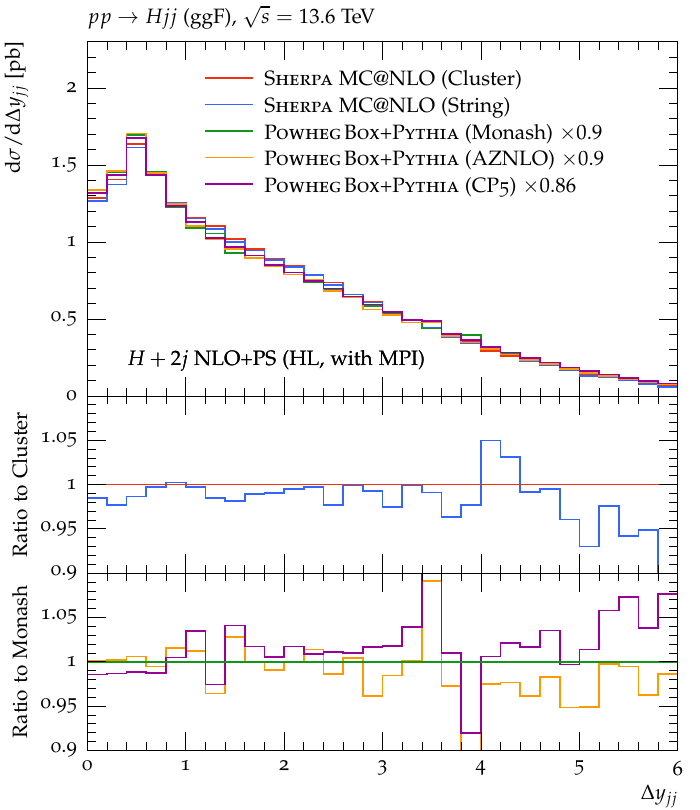}
    \includegraphics[width=0.4\linewidth]{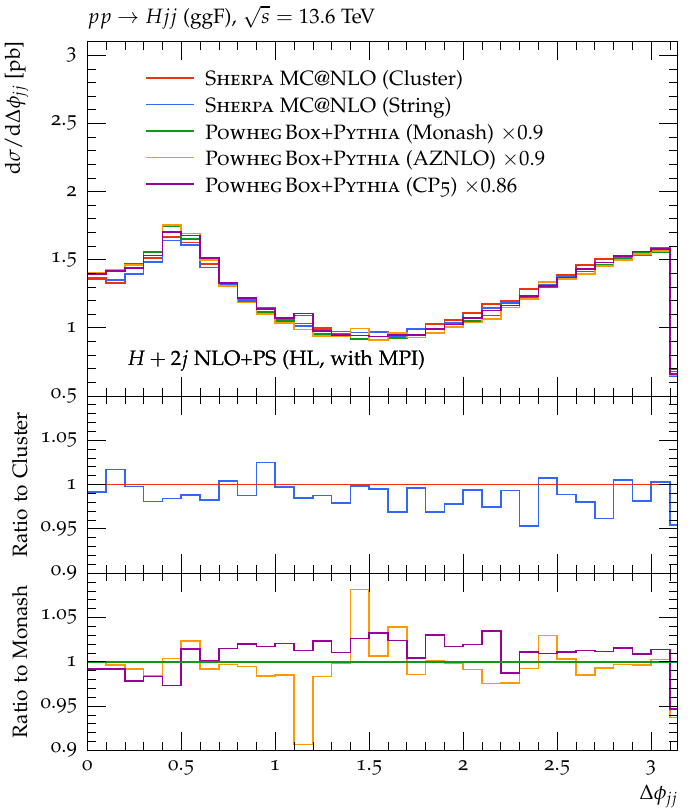}
    \caption{Impact of fragmentation-model variations in \sherpa and tune variations in \pythia in the fiducial $H+2j$ phase space for the Higgs transverse momentum $p_{\mathrm{T},H}$ (top left), the dijet invariant mass (top right), the dijet rapidity separation $\Delta y_{jj}$ (bottom left), and the dijet azimuthal angle separation $\Delta\phi_{jj}$ (bottom right).
    Scaling factors are derived from the \sherpa MC@NLO Higgs rapidity distribution.}
    \label{fig:tuneVariations}
\end{figure}
Fig.~\ref{fig:tuneVariations} shows the comparison of predictions for the same
observables investigates earlier. The differences originating in different
hadronization and underlying event models, as well as the different tunes,
are within 5\%, with the largest discrepancy observed for the CMS C5 tune
compared to the ATLAS AZNLO tune. We note that \herwig predictions cannot be
compared for this setup, because the simulation of the underlying event does
not allow a restriction of the hardness of the secondary scatterings.

\subsection{Comparison of inclusive and exclusive setups}
\label{subsec:inc_vs_nlops}
Finally, we contrast inclusive and exclusive predictions in the
fiducial $H+2j$ phase space. Specifically, we consider the
\sherpa MEPS@NLO results with the \minnlops+~\pythia and \minnlops+~\herwig predictions
as prototypes of inclusive results and compare them to the $H+2j$ NLO predictions
obtained with \sherpa MC@NLO, \powhegbox+~\pythia, and \powhegbox+~\herwig. 
Caution is advised in interpreting this comparison, as the \minnlops predictions
only achieve formal LO accuracy for $H+2j$ configurations {(and have been designed for matching with transverse-momentum ordered showers)},
while all other setups, including the \sherpa MEPS@NLO prediction,
are NLO precise with leading logarithmic accuracy on the parton shower side.

\begin{figure}[t]
    \centering
    \includegraphics[width=0.32\linewidth]{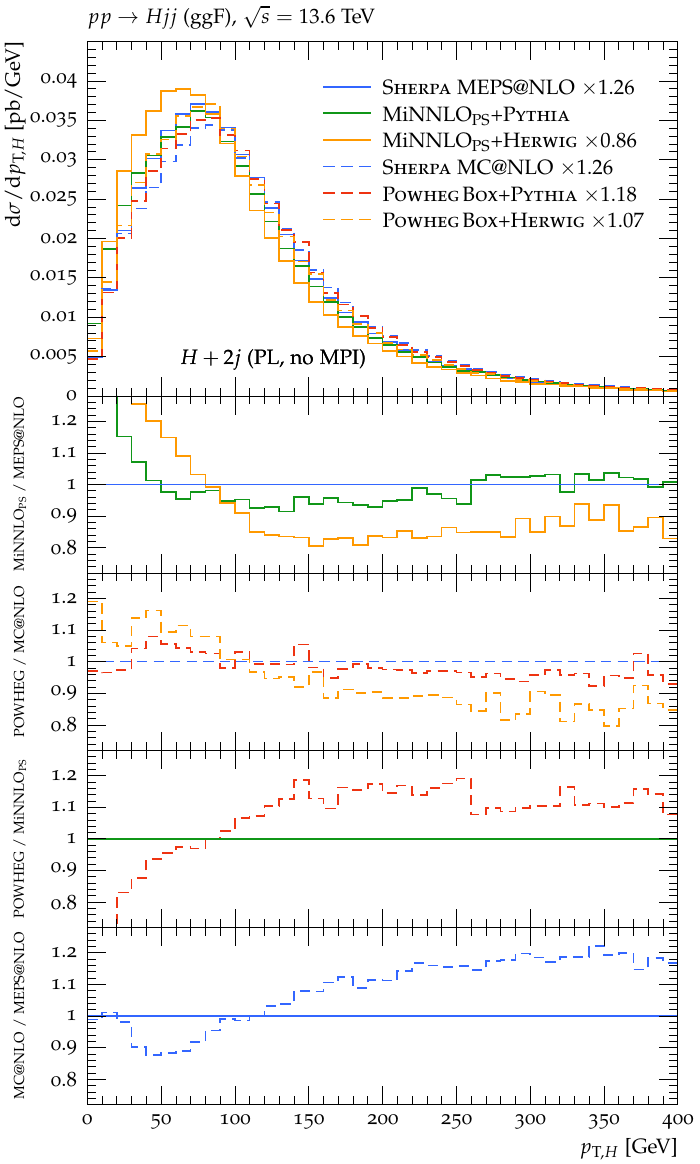}
    \includegraphics[width=0.32\linewidth]{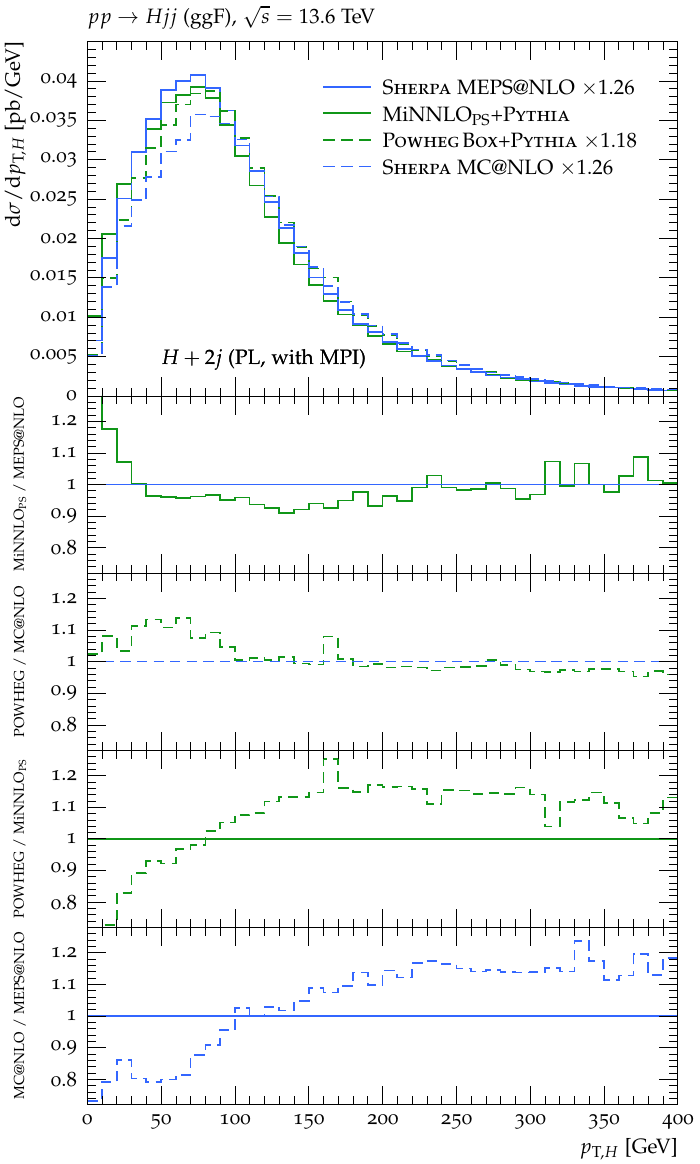}
    \includegraphics[width=0.32\linewidth]{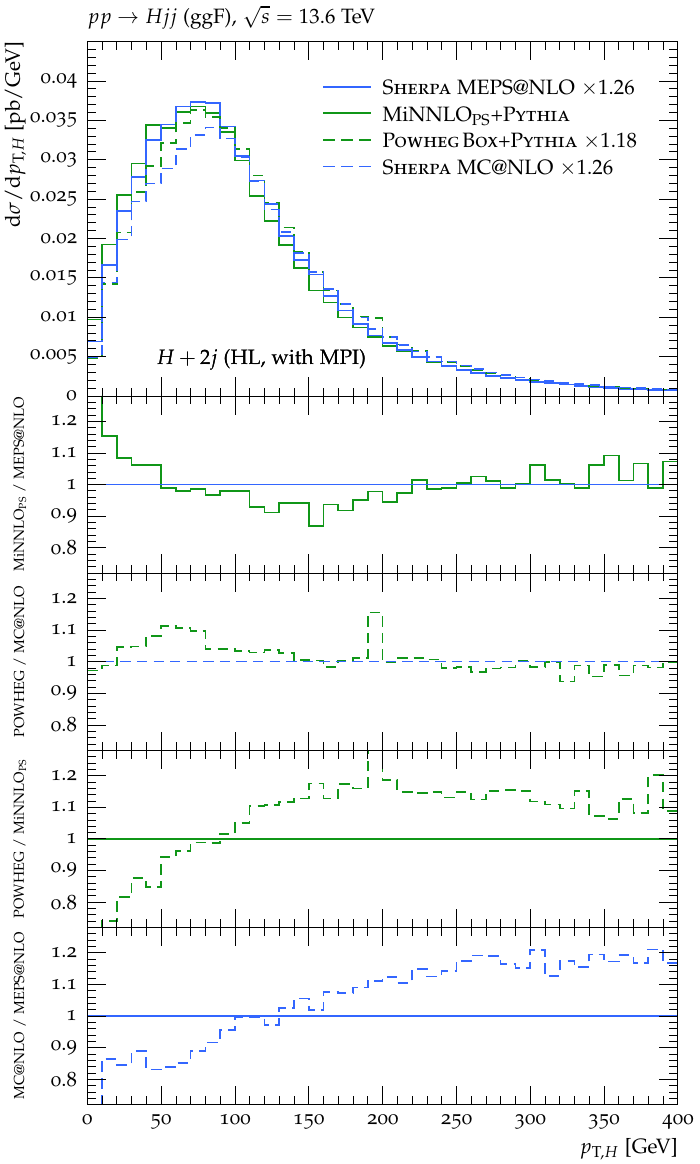}
    \caption{Comparison of the inclusive MEPS@NLO and \minnlops predictions to the matched $H+2j$ NLO+PS predictions in the fiducial $H+2j$ phase space for the Higgs transverse momentum $p_{\mathrm{T},H}$ at parton level (left), parton level with MPI (middle), and hadron level (right). 
    Scaling factors are derived from the \minnlops+~\pythia Higgs rapidity distribution.}
    \label{fig:incVsNLOPSpTH}
\end{figure}
\begin{figure}[t]
    \centering
    \includegraphics[width=0.32\linewidth]{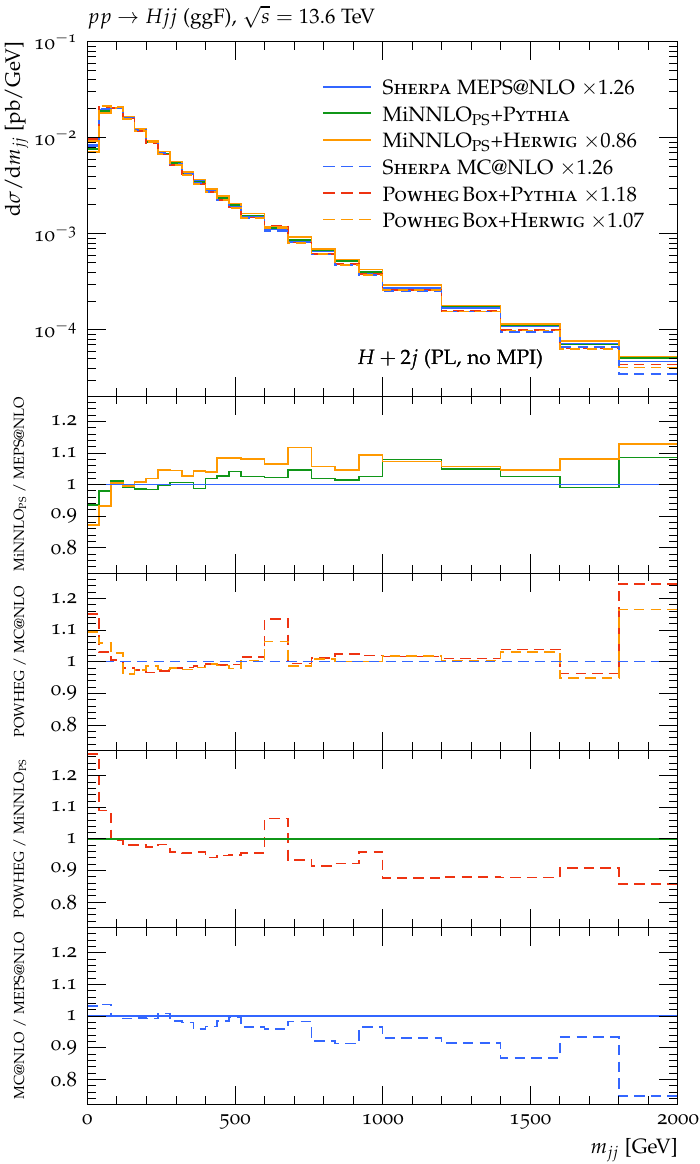}
    \includegraphics[width=0.32\linewidth]{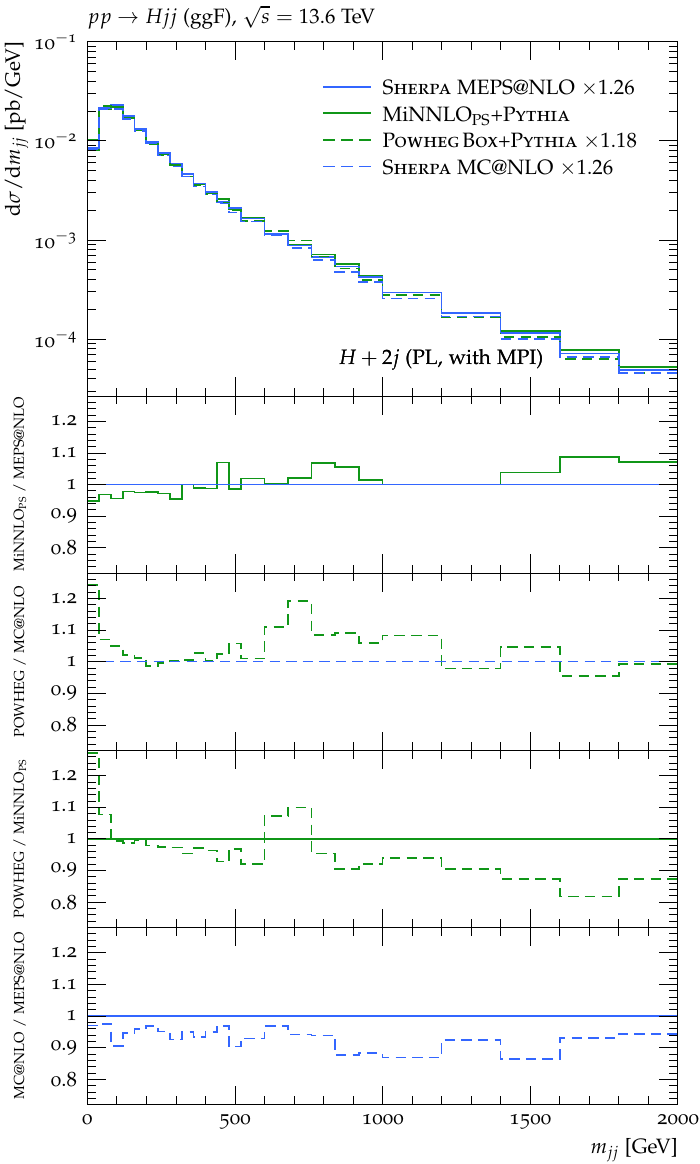}
    \includegraphics[width=0.32\linewidth]{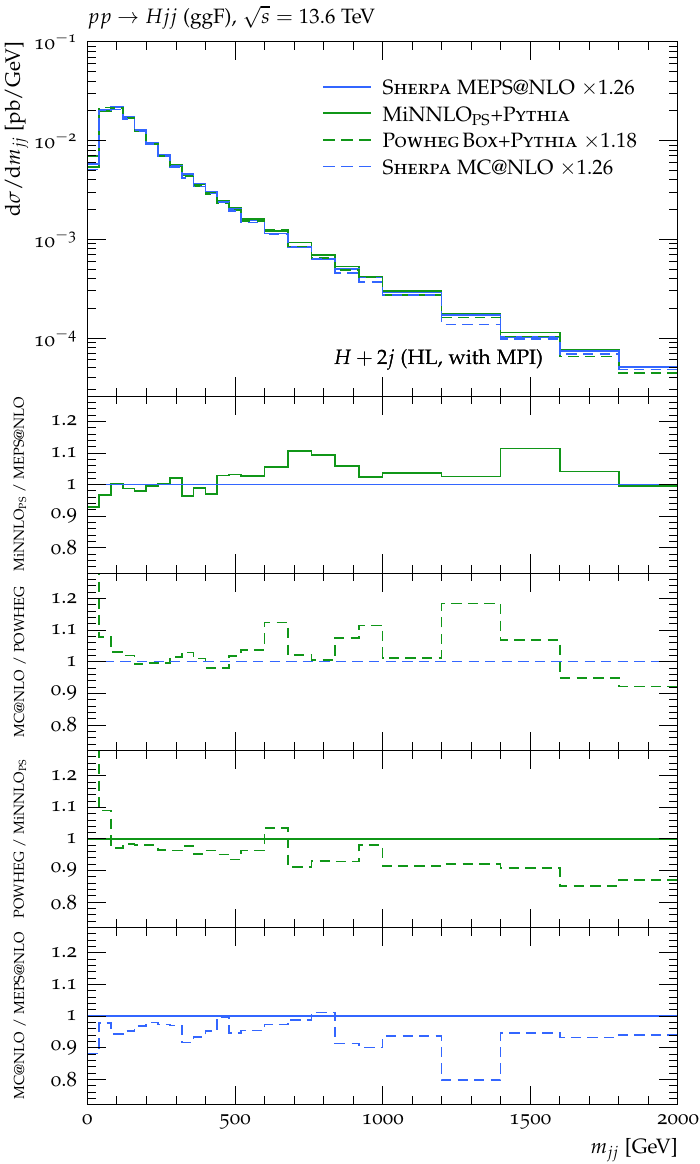}
    \caption{Comparison of the inclusive MEPS@NLO and \minnlops predictions to the matched $H+2j$ NLO+PS predictions in the fiducial $H+2j$ phase space for the dijet invariant mass $m_{jj}$ at parton level (left), parton level with MPI (middle), and hadron level (right). 
    Scaling factors are derived from the \minnlops+\pythia Higgs rapidity distribution.}
    \label{fig:incVsNLOPSmjj}
\end{figure}
\begin{figure}[t]
    \centering
    \includegraphics[width=0.32\linewidth]{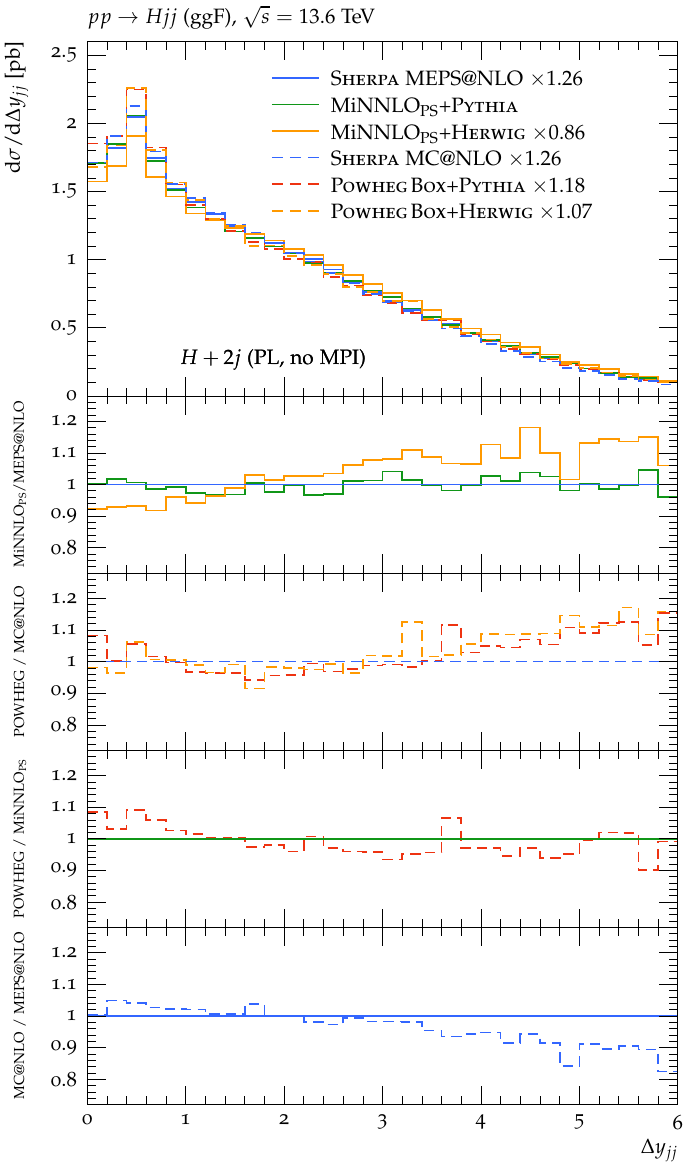}
    \includegraphics[width=0.32\linewidth]{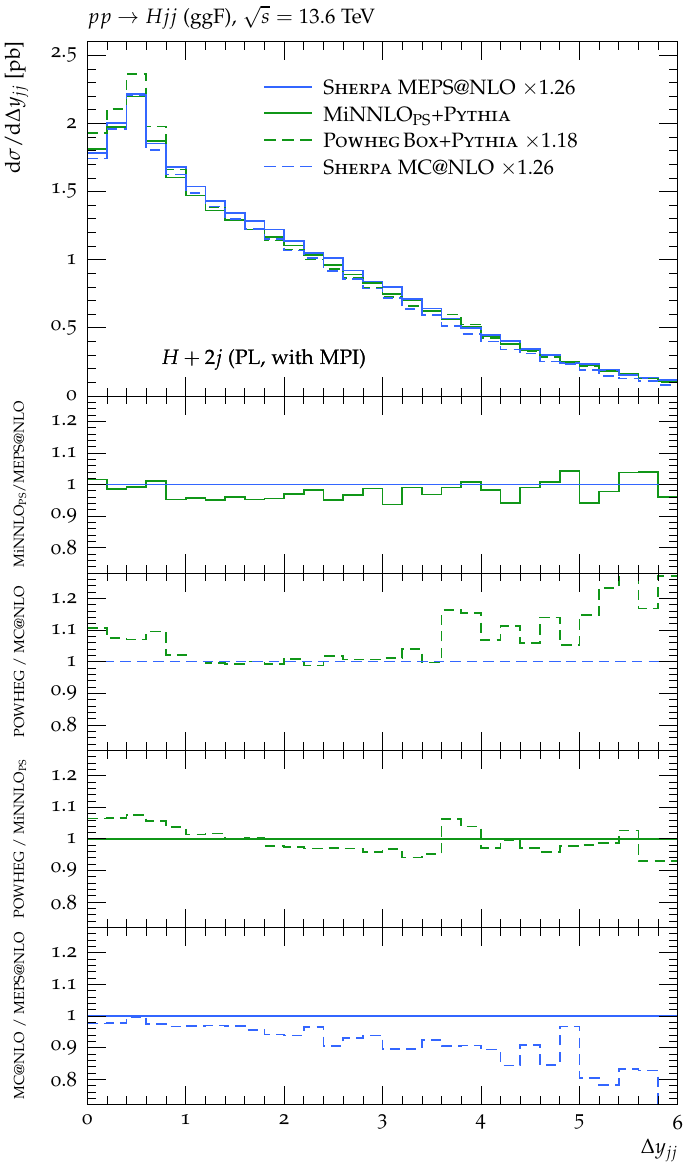}
    \includegraphics[width=0.32\linewidth]{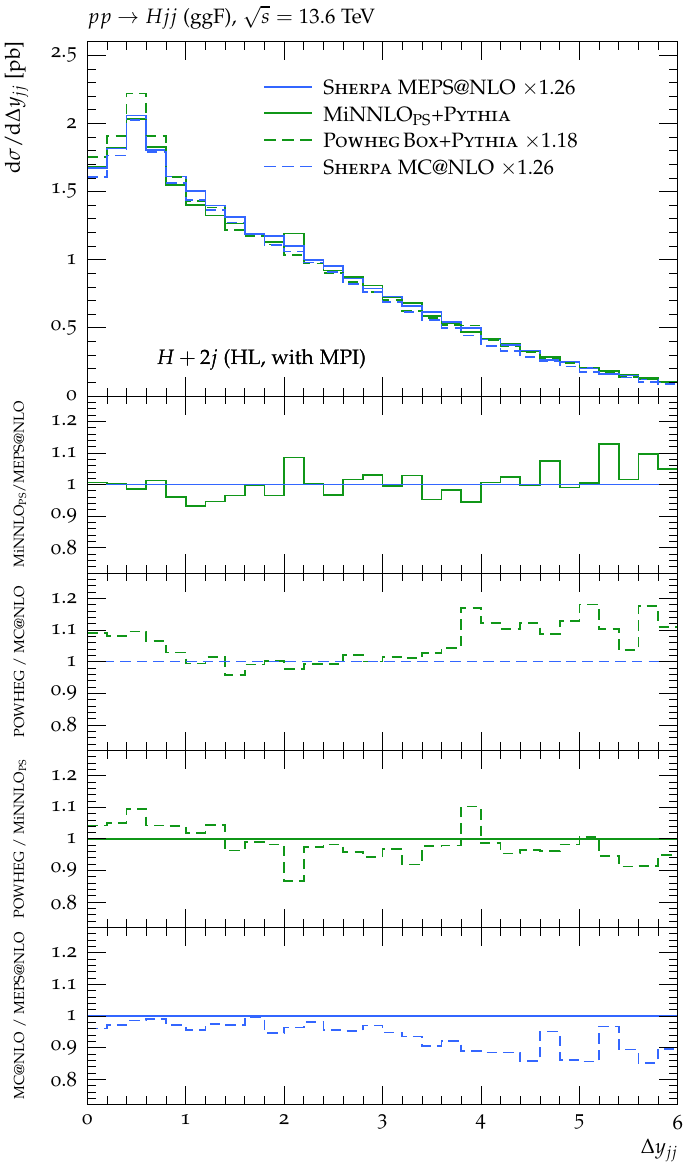}
    \caption{Comparison of the inclusive MEPS@NLO and \minnlops predictions to the matched $H+2j$ NLO+PS predictions in the fiducial $H+2j$ phase space for the dijet rapidity separation $\Delta y_{jj}$ at parton level (left), parton level with MPI (middle), and hadron level (right). 
    Scaling factors are derived from the \minnlops{}+\pythia Higgs rapidity distribution.}
    \label{fig:incVsNLOPSdeltayjj}
\end{figure}
\begin{figure}[t]
    \centering
    \includegraphics[width=0.32\linewidth]{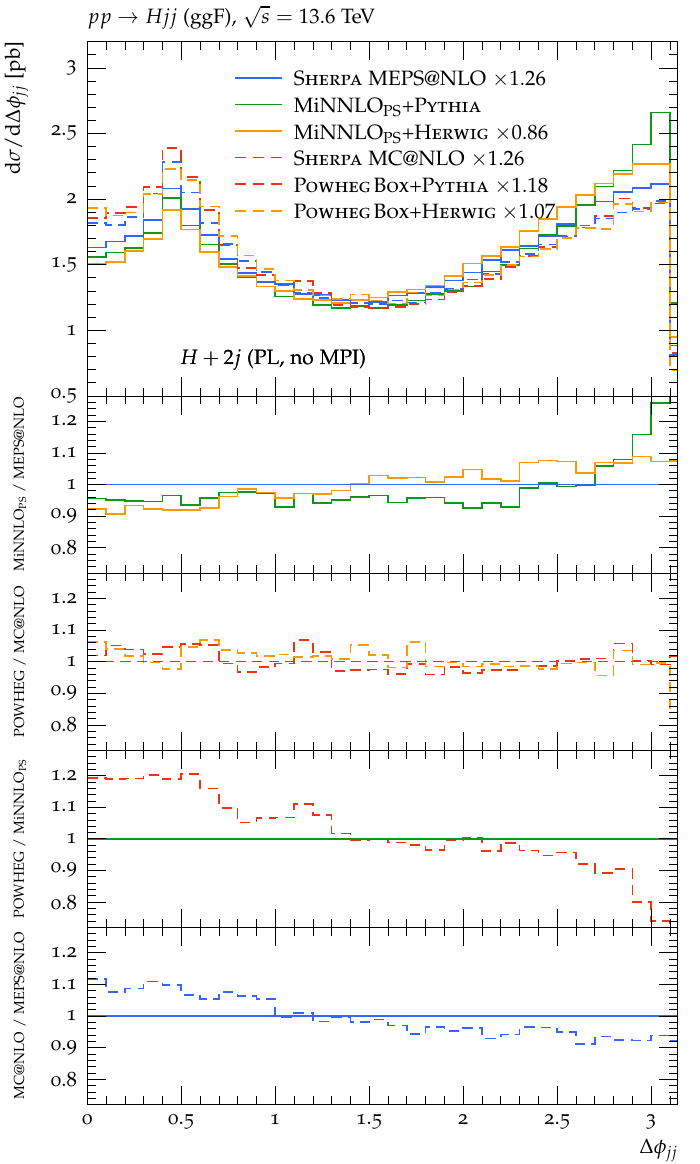}
    \includegraphics[width=0.32\linewidth]{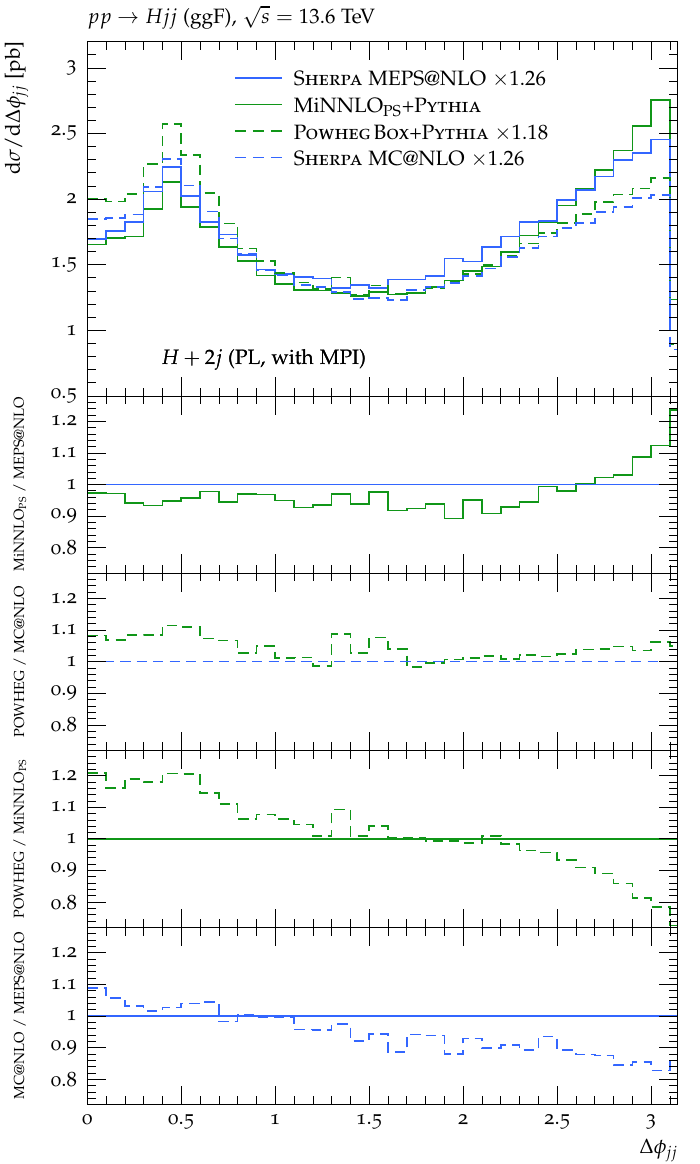}
    \includegraphics[width=0.32\linewidth]{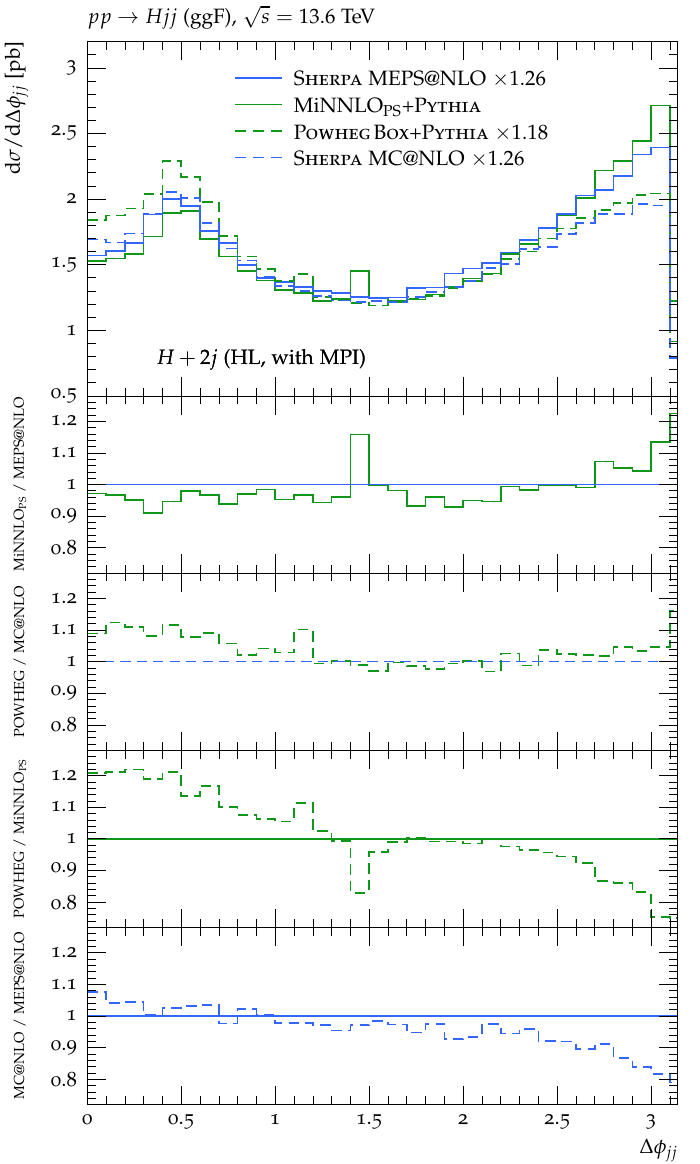}
    \caption{Comparison of the inclusive MEPS@NLO and \minnlops predictions to the matched $H+2j$ NLO+PS predictions in the fiducial $H+2j$ phase space for dijet azimuthal separation $\Delta\phi_{jj}$ at parton level (left), parton level with MPI (middle), and hadron level (right). 
    Scaling factors are derived from the \minnlops+\pythia Higgs rapidity distribution.}
    \label{fig:incVsNLOPSphijj}
\end{figure}
Predictions for {$p_{\mathrm{T},\mathrm{H}}$, $m_{jj}$, $\Delta y_{jj}$, and $\Delta \phi_{jj}$}
are shown in Figs.~\ref{fig:incVsNLOPSpTH}-\ref{fig:incVsNLOPSphijj}. There is some
difference between modeling higher jet multiplicities at leading order (\sherpa MEPS@NLO)
and treating them as part of the inclusive $H+2j$ calculation (\sherpa MC@NLO, \powhegbox+~\pythia, \powhegbox+~\herwig),
in particular for observables that are sensitive to additional hard jets,
such as $p_{\mathrm{T},\mathrm{H}}$. For variables such as $\Delta \phi_{jj}$, and
$\Delta y_{jj}$, differences are smaller.

Perhaps the most striking feature is observed in the comparison of $\Delta \phi_{jj}$,
shown in Fig.~\ref{fig:incVsNLOPSphijj}. There is a marked shape difference between
\minnlops and MEPS@NLO in the region around $\Delta\phi_{jj}\approx\pi$, a region
important for studies of CP-violation, for both VBF Higgs and ggF $\text{Higgs} + 2 \text{jet}$ production.
This difference is amplified at the parton level.
Moreover, there is a dramatic shape difference between \minnlops+~\herwig and
\minnlops+~\pythia, with \minnlops+~\herwig indicating a similar behavior around
$\Delta\phi_{jj}\approx\pi$ as the NLO matched predictions.
As a consequence of these observations, we conclude that the \minnlops simulation
has certain deficiencies when it comes to predicting the azimuthal correlations
between the leading jets. This adds to the problems of describing the leading jet
transverse momentum spectra observed in Fig.~\ref{fig:jetSpectra}.
On the other hand, we observe that all calculations that are NLO precise
for the $H+2j$ final state are consistent in their predictions of the
$\Delta\phi_{jj}$ spectrum. This includes in particular the MC@NLO predictions
from \sherpa, the \powhegbox+~\pythia and \powhegbox+~\herwig predictions, and the
inclusive MEPS@NLO predictions from Sherpa. We have verified that the MEPS@NLO
results also agree with LO merged results from \sherpa, and that the latter are
stable with respect to the number of jets included in the merging, as long as a
third jet is included. 

\section{The transverse momentum of the Higgs-plus-dijet system}
\label{sec:higgspt}
In this section we discuss the transverse momentum spectrum of the Higgs boson
and the two leading jets. This observable is predicted at leading order precision
in the NLO accurate simulations of the $H+2j$ final state employed in this paper.
It probes both the real radiation pattern in the parton-shower approximation,
and the effect of the choice of renormalization and factorization scale.
It is therefore an excellent gauge of the residual differences between the
various event generator predictions at the parton level.

\begin{figure}[t]
    \centering
    \includegraphics[width=0.49\linewidth,valign=t]{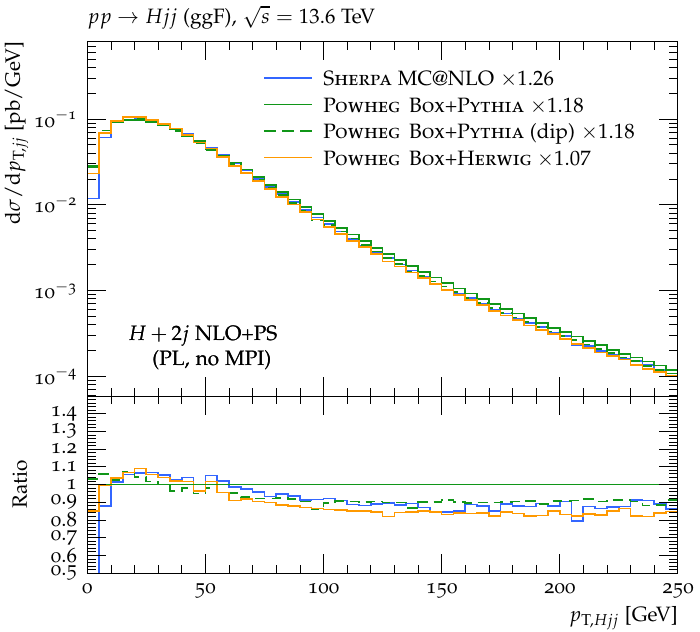}
    \caption{The transverse momentum spectrum of the $H+2j$ system.
    Comparison of the exclusive $H+2j$ NLO+PS setups at parton level without MPI.}
    \label{fig:pTHjj}
\end{figure}
Figure~\ref{fig:pTHjj} shows the results for NLO matched calculations.
The agreement between most of the predictions is at the level of 10\%, which
is well within the range expected from renormalization scale variations
for this complex final state. The largest difference is observed between the
``simple shower'' result from \powhegbox+~\pythia and the prediction from
\powhegbox+~\herwig. In this case, the deviation is about 20\%, which can
still be considered a good agreement, given the complexity of the $H+2j$
Born process.

If an experimental analysis depends significantly on the precise prediction
given by an event generator for the transverse momentum spectrum of the
$H+2j$ system, an improved simulation would be required for this observable.
In particular, the result should then be stabilized by including the $H+3j$ final
state at NLO precision through multi-jet merging. While this would in principle
be possible with the tools employed in this publication, it would require
an outsized investment into computing time, and we therefore postpone the
corresponding study to a future publication.

\section{Cross section in cut based analyses}
\label{sec:cutflow}
In previous sections, we have been comparing relatively inclusive distributions generated by the different Monte Carlos/fixed order programs. The determination of a VBF signal/background region relies on cuts on multiple variables, so in this section we describe a cutflow that examines different kinematic regions, some of which are related to the STXS analyses. 

\begin{figure}[t]
    \centering
    \includegraphics[width=0.49\linewidth]{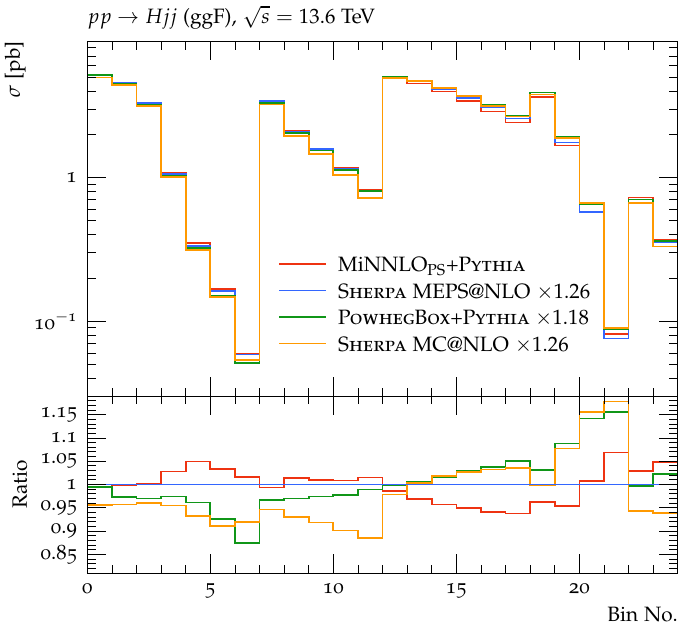}
    \includegraphics[width=0.49\linewidth]{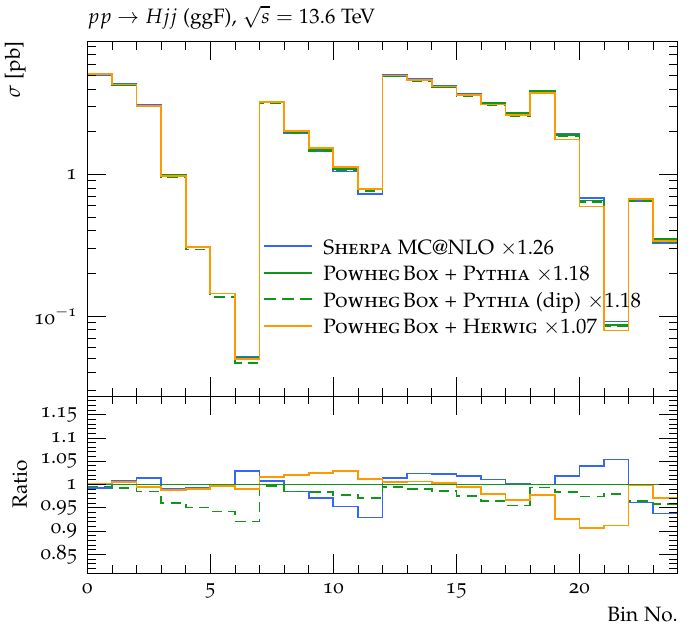}
    \caption{Cutflow for inclusive and exclusive event-generation modes at hadron level with MPI (left), and for different $H+2j$ NLO+PS setups at parton level (right).
    }
    \label{fig:cutflow}
\end{figure}
In Fig.~\ref{fig:cutflow}, we show the cross sections for a number of predictions which are NLO accurate in the two jet phase space, compared to the prediction from \minnlops. Bin 0 shows the inclusive two-jet prediction, satisfying the rapidity and transverse-momentum cuts.%
\footnote{The \sherpa MC@NLO prediction in Fig.~\ref{fig:cutflow} left does not exactly reproduce the inclusive $H+2j$ rate of the
parton-level result in Fig.~\ref{fig:cutflow} right (Bin 0 of the histogram). This is due to the generation cuts and the very loose jet requirements,
which result in a larger effect from jet production through multiple parton scatterings than in \sherpa MEPS@NLO and \powhegbox{}+\pythia.
A more inclusive simulation (such as \sherpa MEPS@NLO) is needed if one wishes to resolve the $H+2j$ final state in this region of the phase space.} 
The left panel shows the results at the hadron level with MPI, and the right panel for different NLO+PS setups at the parton level. All predictions agree well, as expected given the normalization conditions described earlier.  In bins 1-6, minimum cuts are placed on the dijet mass of 60, 120, 350, 700, 1000 and 1500 GeV.  There is reasonable agreement with \minnlops (on the order of 10\%), but \minnlops predicts higher invariant mass final states than the other simulations. In bins 7-11, dijet rapidity separation cuts of 1, 2.0, 2.5, 3.0, 3.5 are applied. There is good agreement among the predictions, although \sherpa MC@NLO predicts smaller rapidity separations than the other simulations. 

Bins 12-17 examine the impact of cuts on the combined Higgs plus dijet transverse momentum, $p_{T,Hjj}$ of 5, 10, 15, 20, 25, 30 GeV. The 25 GeV cuts is the default used in the simplified template cross section analysis. There is good agreement for the 5 GeV cut, as expected since this retains the bulk of the cross section. The \minnlops prediction shows a different slope than the other predictions, leading to a more than 10\% lower cross section for \minnlops for $p_{\mathrm{T},\mathrm{H}jj}$ greater than 25 GeV.
Bins 18-21 examine the impact of cuts on the Higgs boson transverse momentum of 60, 120, 200, and 400 GeV. The differences can reach up to 15\%. 

Finally, in bin 22 we impose a loose VBF cut ($m_{jj}>400$~GeV and $\Delta y_{jj}>2.8$), and in bin 23, we impose a tight VBF cut ($m_{jj}>600$~GeV and $\Delta y_{jj}>3.5$). The Monte-Carlo predictions agree to within 10\% on the rate for these two bins. 

\section{Conclusions}
\label{sec:conclusions}
In this study, we have compared theory predictions for the irreducible background to
Higgs production in vector boson fusion from gluon-gluon fusion. We began with the
prevailing simulation tools used by ATLAS and CMS, which are based on NNLO precise
calculations for the inclusive Higgs boson production process, matched to a parton shower.
These calculations achieve LO precision for the final states of interest in the VBF
measurements and are therefore below the state of the art. In addition, differences
in excess of 20\% have previously been observed between different modes of running
these simulations. We have pinpointed some of the differences, which have the potential
to affect the determination of the CP properties of the Higgs boson.

We have also constructed a theoretically consistent set of simulation setups for
the general-purpose event generators \herwig, \pythia, and \sherpa, matched to
either the \powhegbox or the \sherpa internal S-MC@NLO simulation of $H+2j$ production
at NLO precision. These tools achieve formal NLO precision for the ggF induced
background to VBF measurements and have been validated against \nnlojet.
Comparisons among these tools show consistent predictions,
at the level of 10\% or less, which have been tested against variations
of the matching scheme, the parton shower implementation, the hadronization model,
the simulation of the underlying event, and the non-perturbative tuning parameters.
In addition, we have compared the matched predictions to leading- and next-to-leading
order multi-jet merged results, which provide an inclusive prediction that (in the case
of NLO merging) is able to achieve NLO precision for the $H+2j$ final state as well.

Overall, we find consistent behavior and agreement of all NLO matched results
at the level of 10\%. This uncertainty is independent of, and in addition to, the
conventional renormalization and factorization scale uncertainties as well as the
PDF uncertainties. It is much smaller than the uncertainties observed in previous
comparisons of simulation tools employed by the experimental analyses, which indicates
that the theory systematics may be poorly understood and overestimated. We note in particular,
that we obtain consistent predictions for the dijet azimuthal angle separation,
which is important for understanding the CP properties of the Higgs boson.

{Our recommendation is therefore to rely on any of the $H+2j$ NLO accurate predictions by default, and to gauge the uncertainty of the result by varying among the other $H+2j$ NLO accurate setups presented in this study.}

\section*{Acknowledgments}\noindent
This manuscript has been authored by Fermi Forward Discovery Group, LLC
under Contract No. 89243024CSC000002 with the U.S.\ Department of Energy,
Office of Science, Office of High Energy Physics and the 
National Science Foundation under grant PHY-2309950. 
This research is supported in part by the National Science Foundation of China (NSFC)
with grants No.12475085 and No.12321005.
This research used resources
of the National Energy Research Scientific Computing Center (NERSC),
a Department of Energy Office of Science User Facility using NERSC award ERCAP0028985.
Part of the computations were carried out on the PLEIADES cluster at the
University of Wuppertal, supported by the Deutsche Forschungsgemeinschaft
(DFG, grant No. INST 218/78-1 FUGG) and the Bundesministerium f{\"u}r Bildung
und Forschung (BMBF). 
Another part of the computations were performed on the HPC cluster PALMA II
of the University of M{\"u}nster, supported by the DFG (INST 211/667-1).

We would like to thank the organizers of the Les Houches Workshop on Physics
at TeV Colliders where this project began, and where much of the work was accomplished. 
We are grateful to Simon Pl{\"a}tzer for help with \herwig, especially the \textsc{Matchbox} module.

\appendix

\section{Additional figures}
\label{sec:additionalFigures}
\begin{figure}[t]
    \centering
    \includegraphics[width=0.425\linewidth]{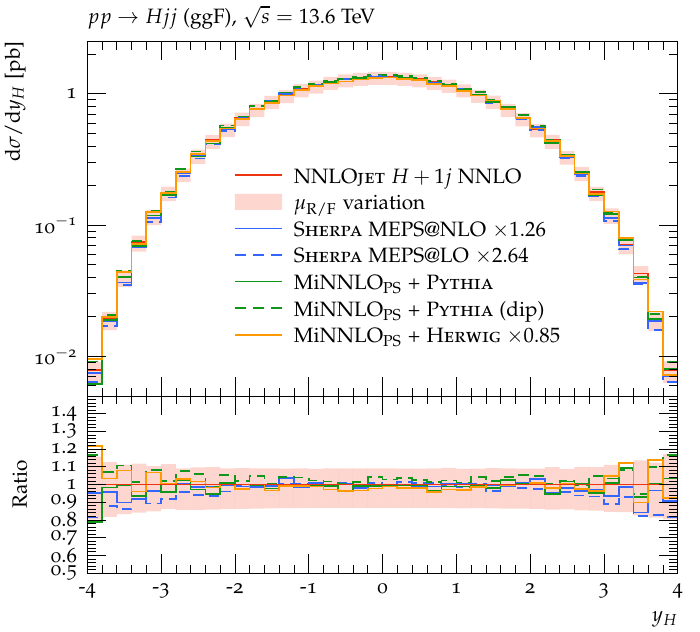}
    \includegraphics[width=0.425\linewidth]{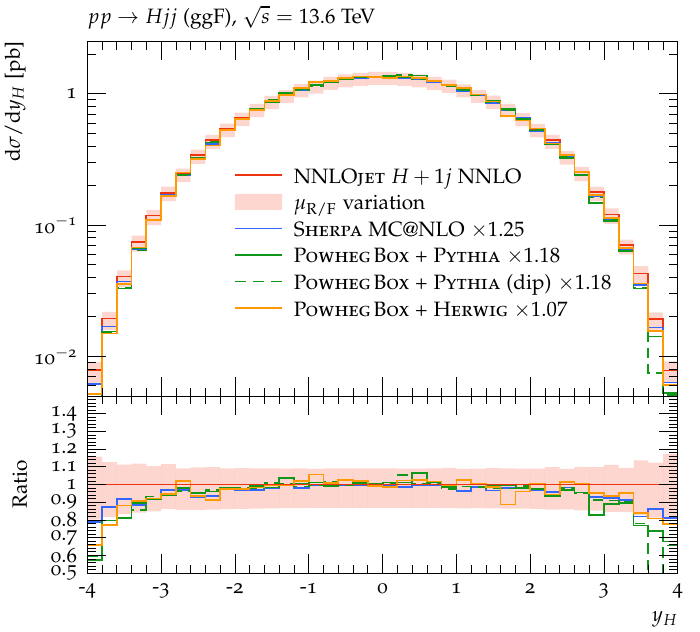}
    \caption{Inclusive (left) and NLO+PS (right) predictions of the Higgs rapidity spectrum in the fiducial $H+2j$ phase space at parton level, scaled to agree with the \nnlojet NLO fixed-order result.}
    \label{fig:higgsRapidity}
\end{figure}

Here, we collect two additional figures not shown in the main text.
Figure~\ref{fig:higgsRapidity}, shows the normalization of the inclusive and NLO+PS predictions to the NLO Higgs rapidity spectrum.

\section{\powhegbox predictions with MiNLO scale setting}
\label{sec:minloPredictions}
\begin{figure}[t]
    \centering
    \includegraphics[width=0.4\linewidth]{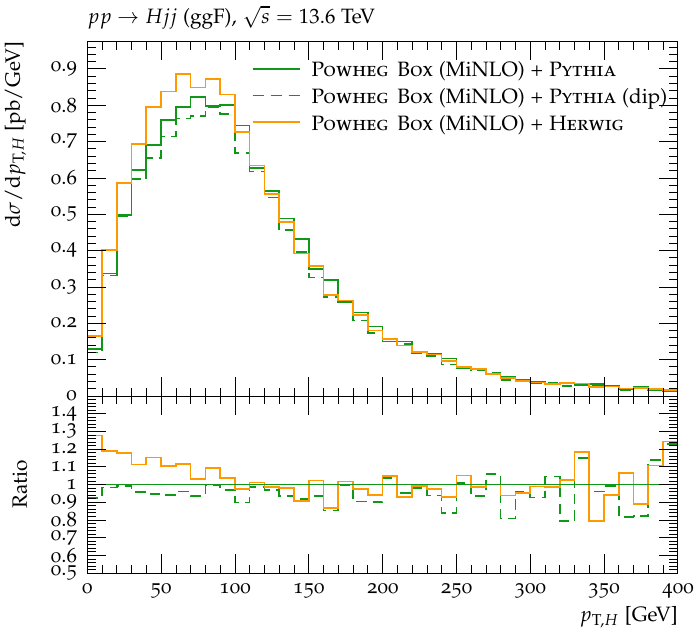}
    \includegraphics[width=0.4\linewidth]{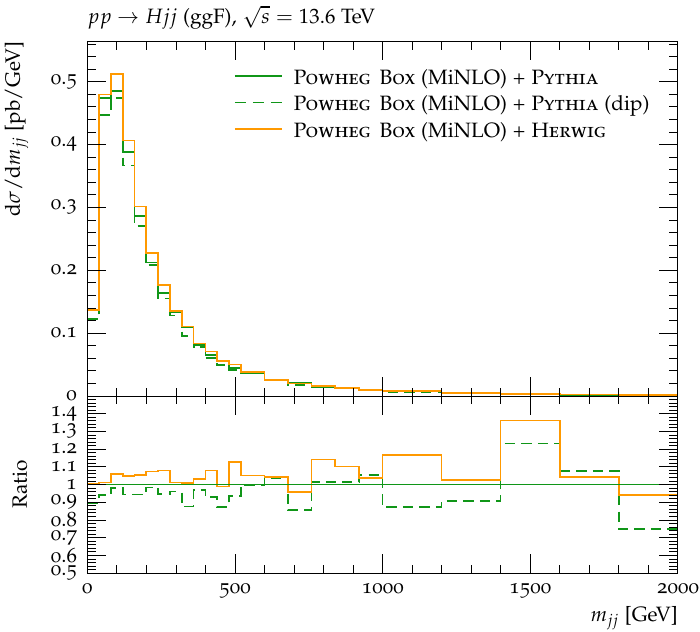}
    \includegraphics[width=0.4\linewidth]{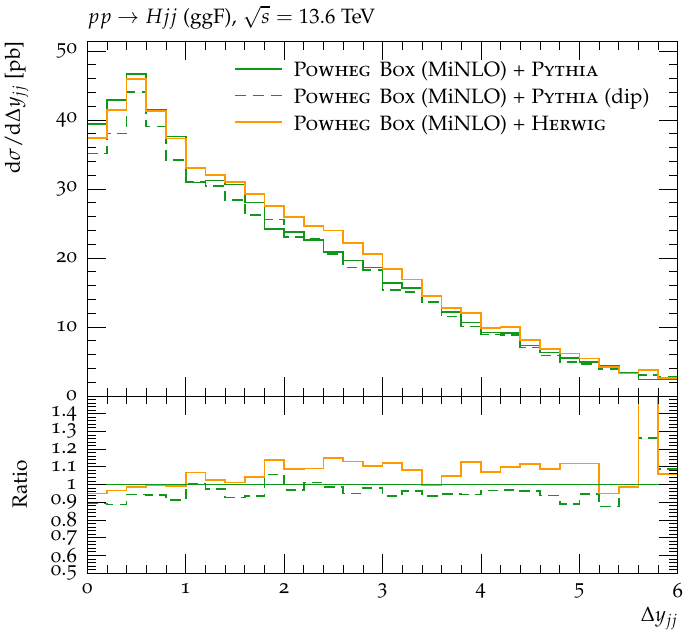}
    \includegraphics[width=0.4\linewidth]{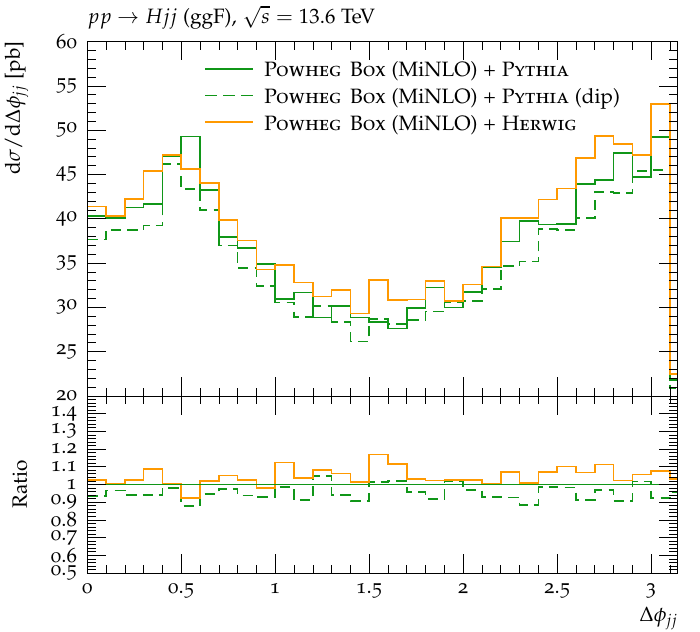}
    \caption{\powhegbox predictions with MiNLO scale setting for the Higgs transverse momentum $p_{\mathrm{T},H}$ (top left), the dijet invariant mass (top right), the dijet rapidity separation $\Delta y_{jj}$ (bottom left), and the dijet azimuthal angle separation $\Delta\phi_{jj}$ (bottom right) in the fiducial $H+2j$ phase space.
    }
    \label{fig:minloPredictions}
\end{figure}

{As an alternative to the Born-suppression method, the $H+2j$ process in \powhegbox can be regulated via the MiNLO prescription.
As MiNLO both introduces Sudakov effects and changes the scale setting of the hard process, it does not agree with the fixed-order benchmarks set by the other event generators.
As such, it is not sensible to include \powhegbox\ MiNLO predictions in the comparisons shown in the main text.
Here, we collect figures comparing \powhegbox~+~\herwig and \powhegbox~+~\pythia, including the default shower and the dipole-recoil option, in the MiNLO prescription at hadron level with MPI.
The results are shown in Fig.~\ref{fig:minloPredictions}.
}

\section{Impact of the EFT approximation}
\label{sec:heftvssm}
\begin{figure}[t]
    \centering
    \includegraphics[width=0.425\linewidth]{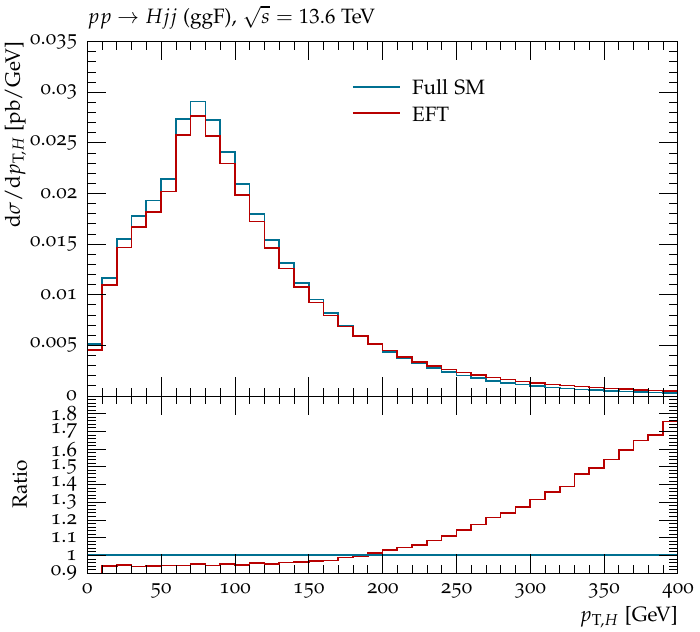}
    \includegraphics[width=0.425\linewidth]{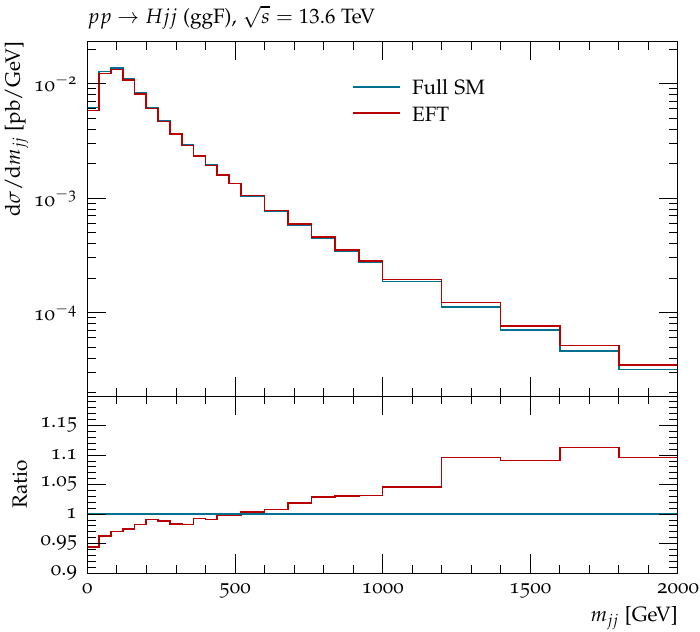}
    \includegraphics[width=0.425\linewidth]{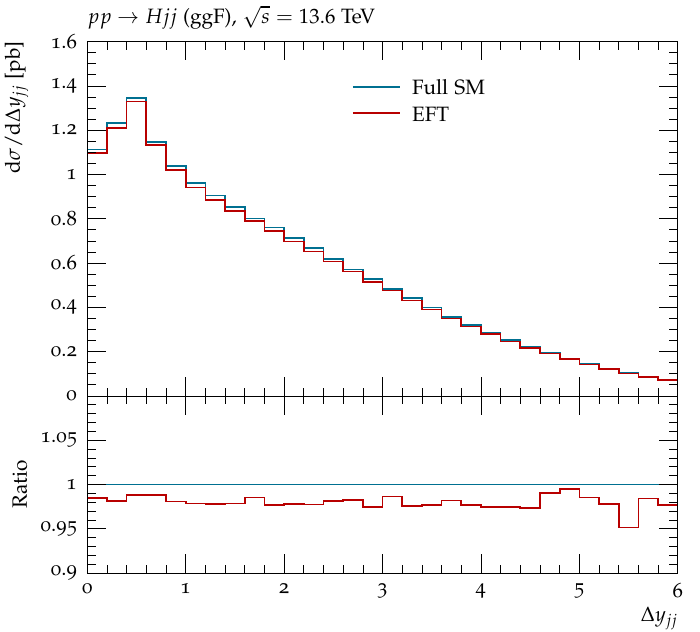}
    \includegraphics[width=0.425\linewidth]{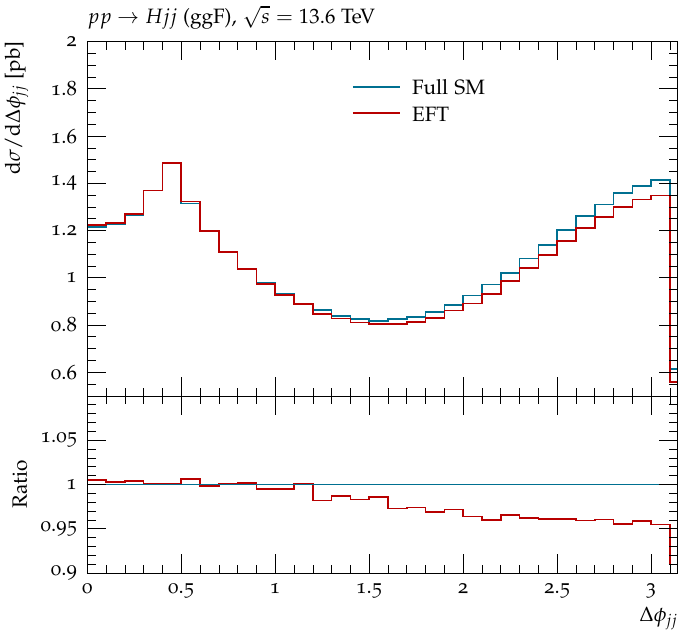}
    \caption{Difference between full SM and HEFT predictions for $H+2j$ production
    at the lowest order in perturbation theory. The plots show the Higgs transverse momentum
    $p_{\mathrm{T},H}$ (top left), the dijet invariant mass (top right),
    the dijet rapidity separation $\Delta y_{jj}$ (bottom left),
    and the dijet azimuthal angle separation $\Delta\phi_{jj}$ (bottom right). 
    The scaling factor is derived from the default SM LO Higgs rapidity distribution.
    }
    \label{fig:eftVsSM}
\end{figure}
It was shown in Ref.~\cite{Greiner:2016awe} that it is important to understand
the impact of the approximations induced by the use of the Higgs effective theory
(HEFT)~\cite{Ellis:1975ap,Wilczek:1977zn,Shifman:1979eb,Ellis:1979jy} when making
predictions for the $H+2j$ final state at NLO precision. In this section,
we therefore investigate the ratio between HEFT and full Standard Model results.
Figure~\ref{fig:eftVsSM} shows the individual results and their ratio, using the
analytic one-loop matrix elements from~\cite{Budge:2020oyl,Campbell:2021vlt}.
The HEFT prediction for the Higgs transverse momentum is approximately 6\% below that
of the Standard Model at low $p_T$, with the ratio rising quickly as the transverse
momentum exceeds the value of the top quark mass, allowing the top quark loop to be
resolved. Due to the requirement of at least two jets in the final state, the
distribution peaks at about 75 GeV. The $\Delta y_{jj}$ ratio, dominated by events
with the Higgs boson transverse momentum below the top threshold, shows a relatively
flat difference (of the order of a few percent) between the standard model and the
HEFT predictions for the entire range. The difference between the HEFT and Standard
Model distributions for the $\Delta\phi_{jj}$ distribution reaches 6\% only at large
$\Delta\phi_{jj}$, corresponding to relatively low values of $p_{\mathrm{T},H}$.
Low values of $\Delta\phi_{jj}$ correspond to the configuration where the Higgs boson
is recoiling against two approximately collinear jets. 

\section{Choice of jet radius}
\label{sec:coneSizeChoice}
In this subsection we investigate the jet radius dependence of the observables introduced in
Sec.~\ref{sec:observables}. From the experimental perspective, the jet radius is a useful tool
to minimize the impact of non-factorizable QCD corrections at hadron colliders, which indicates
that it should be chosen to be as small as possible. On the other hand, a small jet radius leads
to smaller signals and logarithmically enhanced corrections that are due to the vetoed radiation
outside the jet cone. There is thus typically an optimal value, on which experiments decide.
In order to cover the possible choices within ATLAS and CMS, we provide here a study of typical
values, namely $R=0.3$, $R=0.4$, $R=0.6$ and $R=0.6$. We select $R=0.4$ as the central value
for the remainder of the paper.

\begin{figure}[t]
    \centering
    \includegraphics[width=0.425\linewidth]{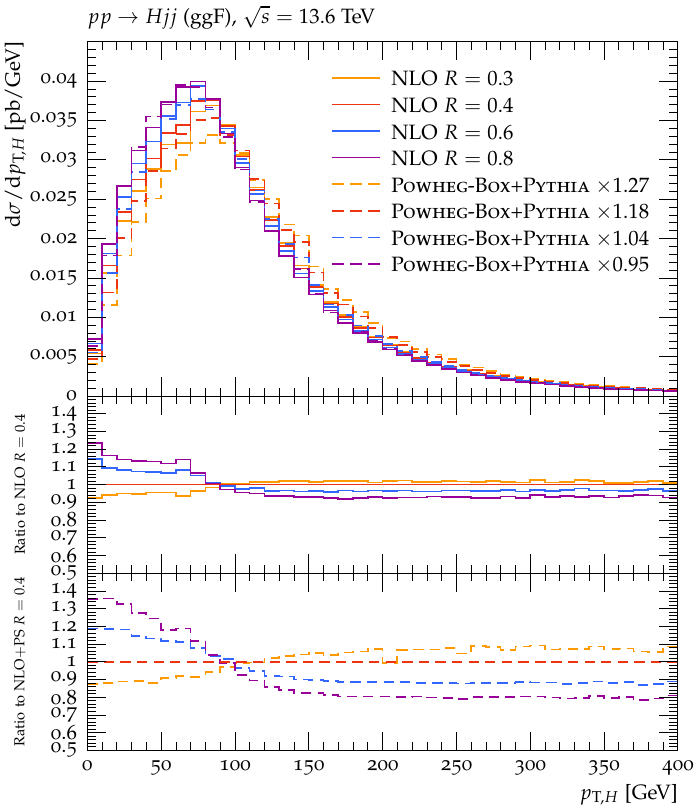}
    \includegraphics[width=0.425\linewidth]{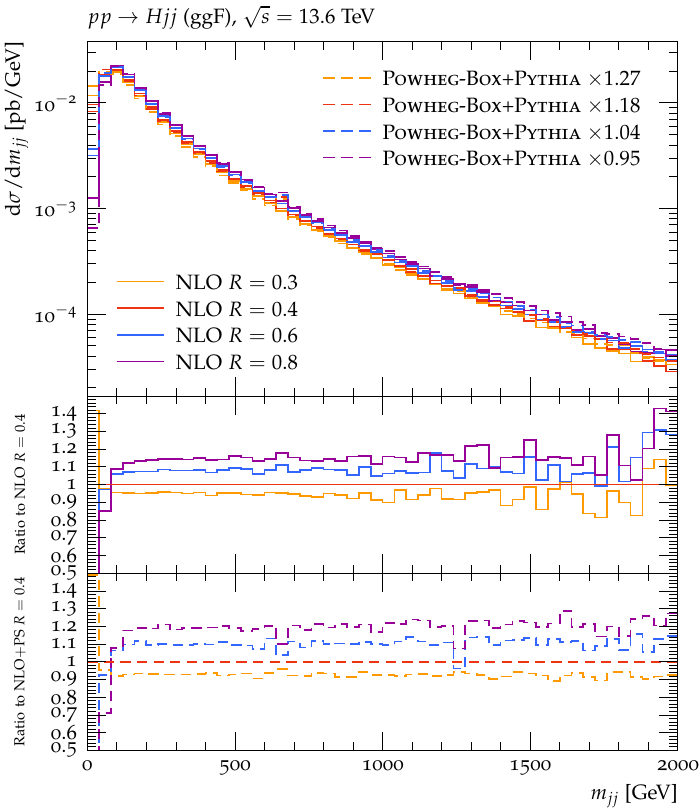}
    \includegraphics[width=0.425\linewidth]{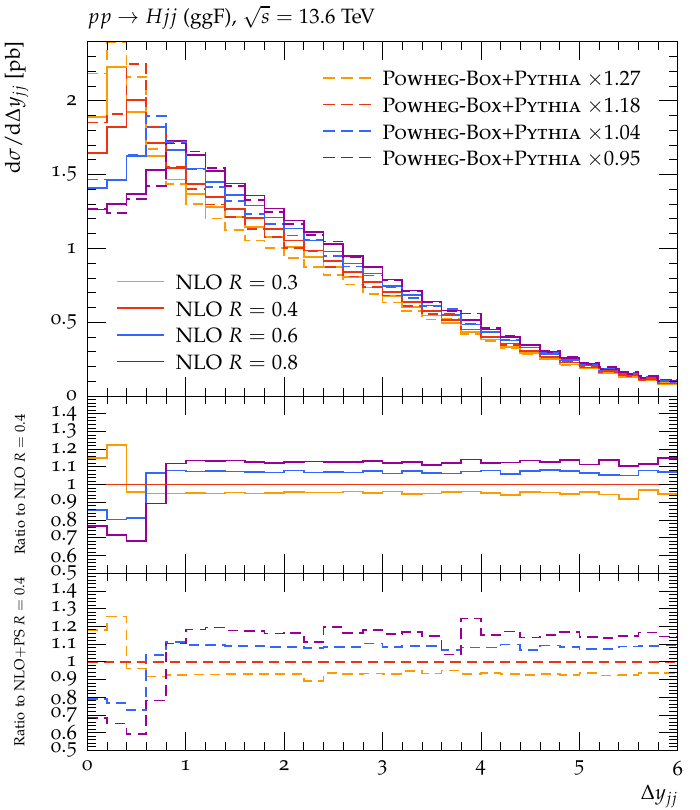}
    \includegraphics[width=0.425\linewidth]{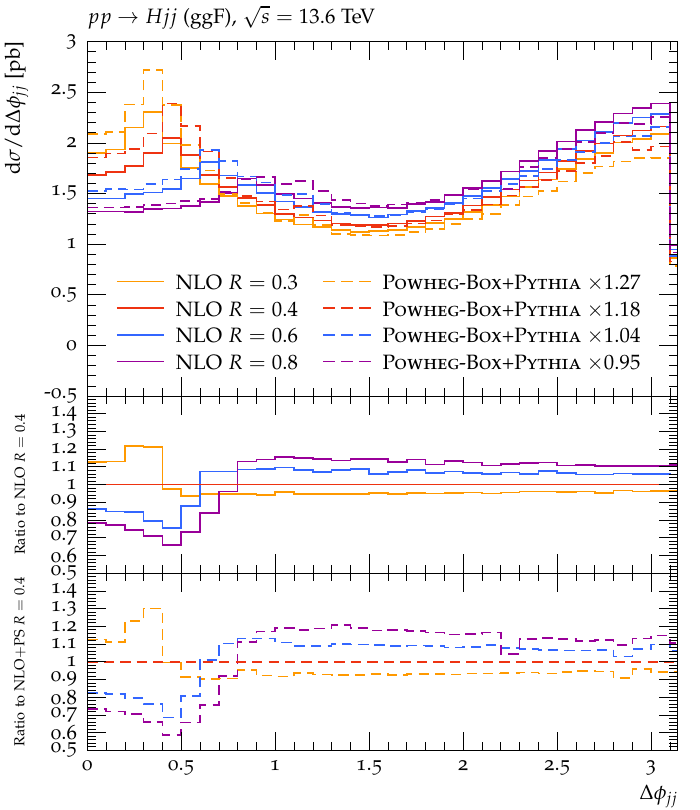}
    \caption{$H+2j$ NLO predictions for different cone-size choices in the fiducial $H+2j$ phase space for the Higgs transverse momentum $p_{\mathrm{T},H}$ (top left), the dijet invariant mass (top right), the dijet rapidity separation $\Delta y_{jj}$ (bottom left), and the dijet azimuthal angle separation $\Delta\phi_{jj}$ (bottom right).}
    \label{fig:coneSizeVariationsFO}
\end{figure}
\begin{figure}[t]
    \centering
    \includegraphics[width=0.425\linewidth]{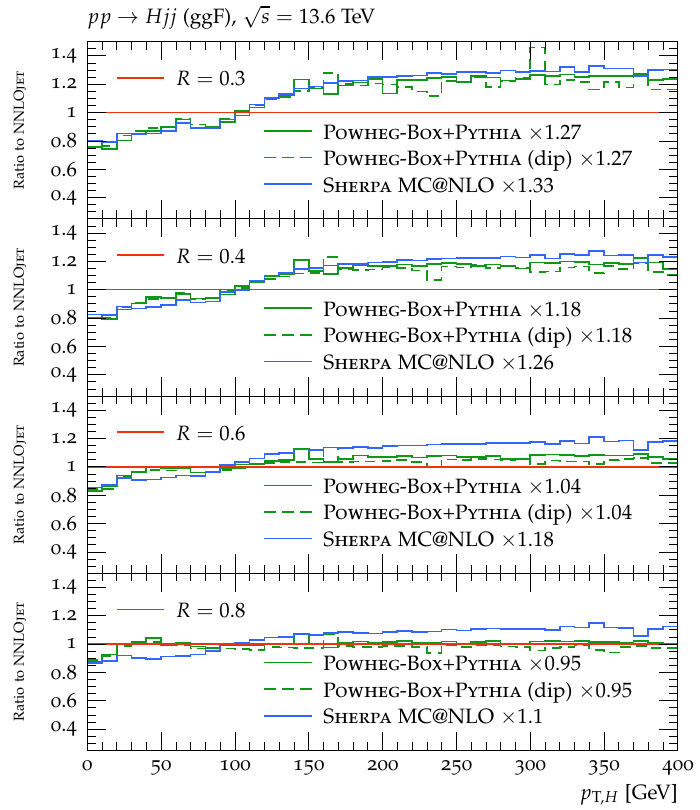}
    \includegraphics[width=0.425\linewidth]{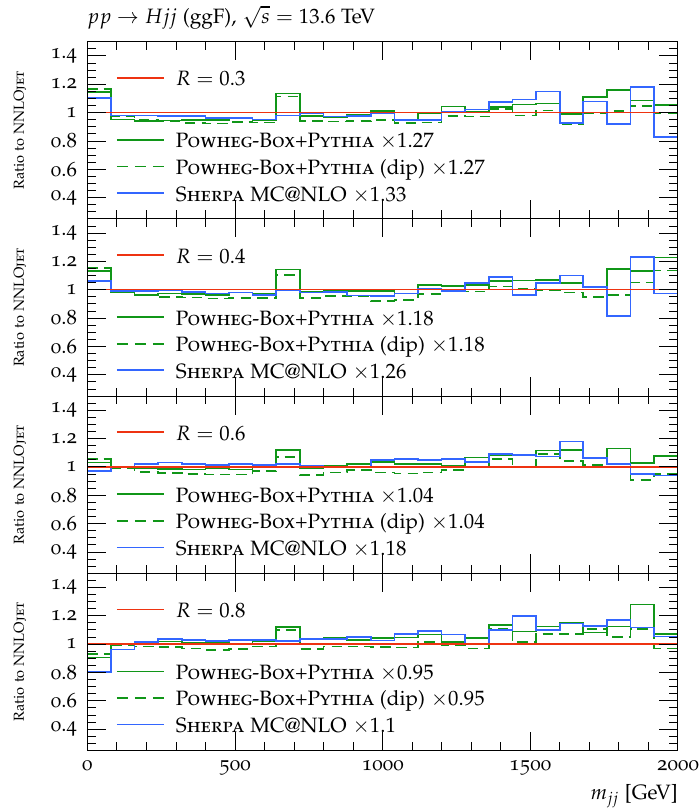}
    \includegraphics[width=0.425\linewidth]{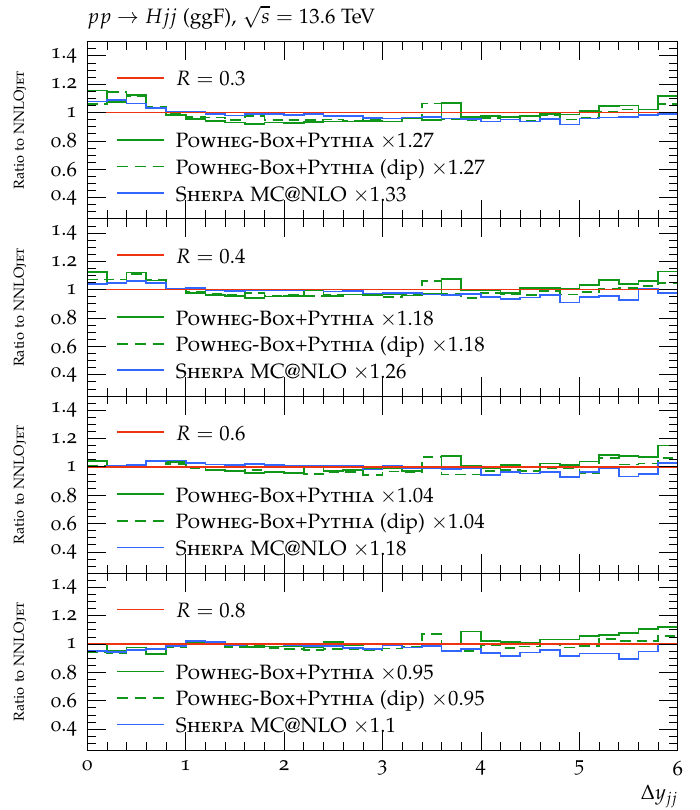}
    \includegraphics[width=0.425\linewidth]{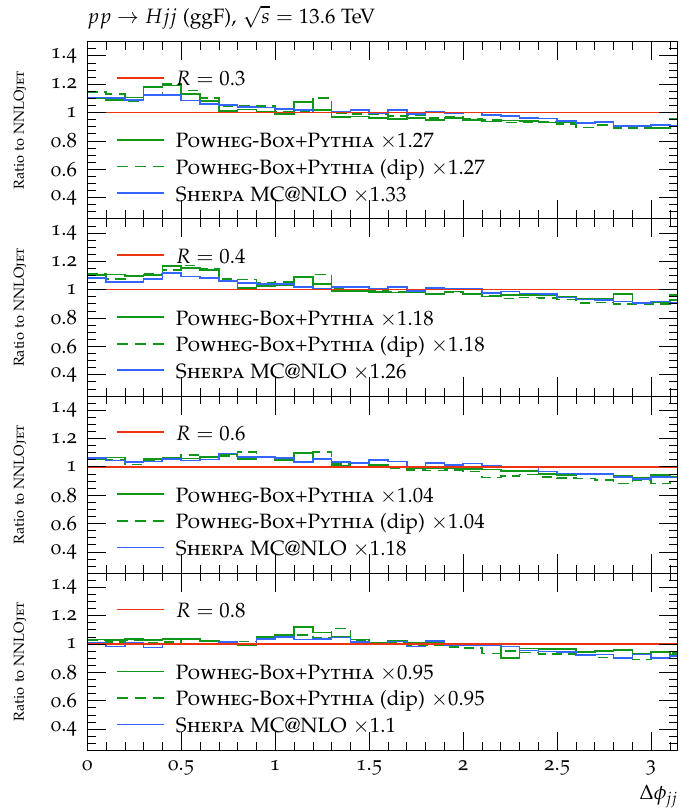}
    \caption{Comparison of different cone-size choices in the \powhegbox+~\pythia $H+2j$ NLO+PS predictions to the \nnlojet $H+2j$ NLO prediction in the fiducial $H+2j$ phase space for the Higgs transverse momentum $p_{\mathrm{T},H}$ (top left), the dijet invariant mass (top right), the dijet rapidity separation $\Delta y_{jj}$ (bottom left), and the dijet azimuthal angle separation $\Delta\phi_{jj}$ (bottom right).
    Scaling factors are derived from the \nnlojet Higgs rapidity distribution, computed at NLO precision.}
    \label{fig:coneSizeVariations_ratio}
\end{figure}
Fig.~\ref{fig:coneSizeVariations_ratio} shows the comparison of \powhegbox+~\pythia $H+2j$ NLO+PS
results to $H+2j$ NLO results obtained with \nnlojet for the different cone-sizes.
We do not observe major differences, but there is generally better agreement between
the next-to-leading order matched predictions and the fixed-order results for larger
cone sizes. The largest differences are observed in the Higgs transverse momentum
spectrum at small values of $R$. In order to be able and judge the variation better
we present in Fig.~\ref{fig:coneSizeVariations_ratio} the ratio of the NLO matched
parton-shower predictions to the fixed-order results for each cone size. We note that
each prediction has been normalized to the $H+2j$ cross section in the fiducial
region, which is inferred from the Higgs boson rapidity spectrum in the $H+2j$
selection. The top left panel of the figure shows again that the most significant
cone size dependent differences appear in the Higgs boson transverse momentum.
This is somewhat expected, because the remaining spectra are less sensitive to
the effects of additional QCD radiation as implemented by the parton shower.

\section{Simulations used in LHC experiments}
\label{sec:atlas_cms_detailed}
In this section we describe the usage of the various existing simulation tools in the experimental collaborations.
This provides the background for the discussion of results in the main text, and the basis for new
recommendations in our conclusions.

\subsection{ATLAS simulations}
\label{subsec:ATLAS}
In the ATLAS experiment, VBF production is simulated with \powhegbox~v2~\cite{Nason:2009ai,Alioli:2010xd,Nason:2004rx,Frixione:2007vw}. Their implementation is based on the corresponding NLO QCD calculations for genuine $W/Z$ vector-boson fusion topologies, using the VBF approximation. Quark--antiquark annihilation and interferences between $t$- and $u$-channel contributions are neglected. The renormalization and factorization scales are set to the $W$ boson mass, and the PDF4LHC\_15nlo PDF set~\cite{Butterworth:2015oua} was used. The matrix elements are matched to the ``simple'' shower algorithm in \pythia~8~\cite{Sjostrand:2014zea}, which employs the AZNLO tune~\cite{ATLAS:2014alx}. 
The dipole-recoil strategy was utilized for the parton shower. The decays of bottom and charm hadrons are implemented by \textsc{EVTGEN}~\cite{Lange:2001uf}. 
The QCD scales $\mu_R$ and $\mu_F$ are varied independently by factors of $0.5$ and $2.0$. The prediction from the \powhegbox sample was normalized to the NNLO cross-section in QCD using the VBF approximation~\cite{Ciccolini:2007jr,Ciccolini:2007ec,Bolzoni:2010xr}. Relative NLO electroweak corrections are also included for the $t$- and $u$-channel contributions considered in the VBF approximation. Alternative predictions are used to estimate the theoretical uncertainty due to variations in parton showering, estimated by showering the same LHE files with the \herwig~7 angular-ordered shower~\cite{Bellm:2017jjp}, using the H7UE set of tuned parameters.

For VBF measurements, the dominant background is typically from gluon-gluon fusion $H+2j$ events. 
In ATLAS, ggF Higgs boson production is simulated at NNLO accuracy in QCD using \powhegbox~v2~\cite{Hamilton:2013fea,Hamilton:2015nsa,Alioli:2010xd,Nason:2004rx,Frixione:2007vw}.
The simulation achieved NNLO QCD accuracy for inclusive $H$ observables by reweighting the Higgs boson
rapidity spectrum in \texttt{HJ}-\texttt{MiNLO}~\cite{Hamilton:2012np,Campbell:2012am,Hamilton:2012rf} to that of \texttt{HNNLO}~\cite{Catani:2007vq}. 
The transverse-momentum spectrum from this MC is found to be compatible with fixed order HNNLO calculation and \texttt{HRes}~2.3~\cite{Bozzi:2005wk,deFlorian:2011xf} performing resummation at NNLL+NNLO.
Top- and bottom-quark mass effects are included up to NLO.
The simulations use the same PDF set as that of VBF, and the predictions are normalized to N3LO in QCD with NLO electroweak corrections~\cite{LHCHiggsCrossSectionWorkingGroup:2016ypw}.

The renormalization and factorization scales are set to half of the Higgs boson mass, and QCD scale uncertainties are estimated using nine-point scale variations of the NLO renormalization and factorization scales and applying the NNLO reweighting to those variations, including up and down variations of $\mu_R=\mu_F$ around the central value for the NNLO part, yielding a total of 27 scale variations~\cite{ATLAS:2023qdv}.
Note that this may lead to an overestimate of the perturbative QCD uncertainty because the scale variations are treated as uncorrelated.
The samples are showered using \pythia~8 with an alternative sample using \herwig~7.

Multiple ATLAS VBF measurements, such as those from the $H\to\gamma\gamma$~\cite{ATLAS:2022tnm}, have leading systematic uncertainties related to Higgs modeling. 
Figure 5b in \cite{ATLAS:2022tnm} shows that parton-shower uncertainties are the leading sources, exceeding even experimental uncertainties for jet calibration.
This uncertainty is estimated by taking the envelope of the differences between the \powhegbox+~\pythia and \powhegbox+~\herwig predictions
and can reach up to 20\%, increasing with jet multiplicities \cite{ATLAS:2022tnm}. The differences are typically on  yields affecting both VBF and ggF backgrounds. They appear in analyses as differences in acceptances, efficiencies, or expected yields. 
A sub-dominant uncertainty comes from QCD variations of the ggF background in the VBF-enriched region, as the QCD accuracy of the NNLO Monte Carlo deteriorates when requiring two jets.

\subsection{CMS simulations}
\label{subsec:CMS}
The CMS Collaboration simulates the VBF Higgs production process at NLO QCD using the \powhegbox~v2~\cite{Nason:2009ai,Alioli:2010xd,Nason:2004rx,Frixione:2007vw} generator.
The dipole-recoil option of the ``simple shower'' algorithm in \pythia~8~\cite{Sjostrand:2014zea} is used to obtain a faithful description of the initial-final color flow that takes into account the color connection between the incoming and the outgoing partons. 
In certain channels, such as the boosted Higgs to $b\bar b$ channel~\cite{HIG-21-020}, the VBF samples are further reweighted to account for NNLO QCD corrections to the $p_T$ spectrum and N$^3$LO QCD corrections to the inclusive cross section~\cite{Cacciari:2015jma,Dreyer:2016oyx}.

In contrast, the ggF Higgs boson production process is simulated using the \texttt{HJ}-\minlo event generator within \powhegbox~\cite{Hamilton:2015nsa, Hamilton:2013fea}, with the Higgs boson mass set to $125~\gev$, including finite top-quark mass effects. 
{Here, \powhegbox~v2 provides the NLO QCD matrix-element calculation matched to a parton shower, while \texttt{HJ}-\minlo specifies the ggF $H{+}$jet setup and MiNLO treatment.}

For channels such as the $H \to ZZ$~\cite{HIG-19-001}, the kinematic features of events produced in ggF with two associated jets are considered. 
The \minlo program at NLO in QCD is utilized for evaluating systematic uncertainties related to the modeling of two associated jets and contamination in the VBF measurement. The two-jet final state is produced with leading order accuracy. 

For the $H\to WW$ decay~\cite{HIG-19-002, HIG-19-009}, the gluon-gluon fusion (ggF) events are simulated at NLO QCD using \powhegbox~v2 and additionally reweighted to match the NNLOPS predictions in the distributions of $p_{T}^{H}$ and $N_{\mathrm{jet}}$.
The ggF sample is normalized to N3LO QCD accuracy and NLO electroweak accuracy.

In the $H\to b\bar{b}$~\cite{HIG-22-009} and $H\to\mu^+\mu^-$~\cite{CMS:2018nak}, the VBF signal is generated using \powhegbox~v2 at NLO QCD accuracy and matched to the ``simple shower'' in \pythia~8, using the dipole-recoil option.
Alternatively, the \powhegbox matrix-element generator interfaced with \herwig~7 is sometimes used for fragmentation and hadronization.
This alternative configuration helps to assess the systematic uncertainty related to the choice of showering and hadronization model.

For parton showering and hadronization, most of the generated samples are interfaced with \pythia~8 using the CP5 tune for the underlying-event description, and the NNPDF3.1 set at NNLO QCD accuracy is used for all processes.
For \herwig~7, the parton shower relies on the so called CH (CMS Herwig) tunes \cite{CMS:2020dqt}, which are based on the NNPDF3.1 NNLO QCD PDF set.

\section{Simulation setups and event samples}
The setups and nominal event samples used in this study are available at \zenodoplotref.\\

\bibliography{main}
\end{document}